 \documentclass[11pt]{article}

 \pdfoutput=1

\usepackage[english]{babel}

\usepackage[round]{natbib} 

\usepackage{graphicx,siunitx,xspace}
\usepackage{amsmath}
\usepackage{amssymb}
\usepackage{hyperref}
\usepackage{enumitem}
\usepackage{color}
\usepackage{enumitem}
\usepackage{float}
\usepackage{amsfonts,bm}
\usepackage{epstopdf}
\usepackage[mathscr]{euscript}

\usepackage{bigints}

  \usepackage[normalem]{ulem}
  \usepackage{textcomp}

\usepackage[utf8]{inputenc}
\usepackage[T1]{fontenc}
\usepackage{pifont} 

\usepackage{chemformula}

\usepackage{float}
\usepackage{ifthen}
\begin{document}


\begin{center}
{\bf \Large
An English translation of the Habilitation Thesis}
\\ \vspace*{2mm}
{\bf \Large
of Max \citet[][]{Planck_Habilitation_thesis_1880}}
\\ \vspace*{2mm}
\hspace*{-0mm}
{\bf \Large 
\underline{\"Uber Gleichgewichtszust\"ande isotroper K\"orper}
\vspace*{2mm}\\ 
\underline{in verschiedenen Temperaturen}}
\\ \vspace*{2mm}
{\bf \Large \underline{On the equilibrium states of isotropic bodies at different temperatures}}
\\ \vspace*{2mm}
{\bf \Large
to provide a readable version of the German content.
}
\\ \vspace*{2mm}
{\bf \large
Translated by Dr.Hab. Pascal Marquet 
}
\\ \vspace*{2mm}
{\bf\bf\color{red}  \large Possible contact at: 
    pascalmarquet@yahoo.com}
    \\
{\bf\bf\color{red} 
    Web Google-sites:
    \url{https://sites.google.com/view/pascal-marquet}
    \\ ArXiv: 
    \url{https://arxiv.org/find/all/1/all:+AND+pascal+marquet/0/1/0/all/0/1}
    \\ Research-Gate:
    \url{https://www.researchgate.net/profile/Pascal-Marquet/research}
}
\\ \vspace*{1mm}
\end{center}

\hspace*{65mm} Version-1 / \today
\vspace*{-3mm}

\begin{center}
--------------------------------------------------- 
\end{center}
\vspace*{-9mm}

  \tableofcontents


\vspace*{2mm} 
\begin{center}
--------------------------------------------------- 
\end{center}
\vspace*{-9mm}

\bibliographystyle{ametsoc2014}
\bibliography{Book_FAQ_Thetas_arXiv}
\vspace*{-3mm}

\begin{center}
--------------------------------------------------- 
\end{center}
\vspace*{-2mm}

Uncertainties/alternatives in the translation are indicated {\color{red}(in red)} with {\it\color{red} italic terms}, together with some additional informations and  footnotes {\color{red}(indicated with {\it P. Marquet)}}.
Moreover, I have added some highlight (shown as \dashuline{\,dashing text}), in particular about the \dashuline{\,integration constant issue for the entropy}.

Do not hesitate to contact me in case of trouble in the English translation from the German text.
\vspace*{-1mm}


\vspace*{-1mm} 
\begin{center}
--------------------------------------------------- 
\end{center}
\vspace*{-2mm}

\newpage

\begin{center}
{\bf \Large \underline{On the equilibrium states of isotropic bodies at different temperatures}}
\end{center}
\vspace*{-4mm}

\begin{center}
{\bf \Large by Max Planck, Munich (1880).}
\end{center}
\vspace*{0mm}

\setcounter{section}{-1}

\vspace*{-4mm}
\section{\underline{Introduction} (p.1-2)}
\vspace*{-2mm}

The question of the conditions of equilibrium of a body subject to certain external circumstances depends on the knowledge of the forces at work within it.

However, these depend not only on the positions which the smallest particles of the body mutually occupy, but also on the temperatures in which they are located, and which generally change with every alteration of state of the body. 
The influence that the displacements of the smallest body particles have on the internal forces is dealt with in detail in the theory of elasticity, while the temperature is usually tacitly assumed to be constant there.

The task of the following treatise is therefore first of all to describe the influence of temperature on the elastic forces inside a body, and the principles of the mechanical theory of heat provide sufficient means for this, without the need to make special assumptions about the molecular nature of the bodies: rather, it is sufficient to assume that these are constantly satisfied by the matter.

We restrict ourselves to the consideration of isotropic bodies, but without regard to their state of aggregation {\it (or composition? or physical state?)}, and assume that the state of an isotropic body of a certain chemical composition and mass is completely determined by its temperature and by the relative position of its elements to each other.

Furthermore, it should be noted that we shall disregard all capillarity effects, and also those forces that act on the mass of the body from a distance, e.g. gravity, because even the latter only has a very minimal influence on the equilibrium conditions inside a body that is not too large.

The necessary conditions of equilibrium, as they result from the use of the circumstance that in a body in equilibrium every element of it, conceived as rigid, is also in equilibrium, are developed in the first section, while the second section contains those conditions which are sufficient for equilibrium, and which naturally include the former, but, as will be shown, are somewhat more extensive.

The results obtained are in complete agreement with those of the mechanical theory of heat and the theory of elasticity, in so far as they touch
on what is known, 
but they will also, especially in the second section, contain some new things, among which a theorem of practical importance, which has not yet become known, may be of particular importance, which, among other things, can be used to calculate the specific heat of a vapour at constant pressure from data which can be determined relatively easily by experiment.

\section{\underline{Necessary equilibrium conditions} (p.3-32)}
\vspace*{-2mm}

Let us imagine any isotropic body (solid, liquid or gaseous) in a state of equilibrium, which we fix as the initial state, 
compared to the changes in state under consideration.

In this initial state, let the body be completely homogeneous and of absolute temperature $T$, and let it be subjected to a positive or negative external pressure which is the same everywhere and acts normally on its surface. This pressure is then kept in equilibrium by internal elastic forces. Let the volume of the unit mass (specific volume) in the initial state be denoted by $v$.

Now let the state of the body change, partly by the action of any external forces, partly 
through the communication or withdrawal
of heat, until a new state of equilibrium is reached, to which the temperature $T'$ and the specific volume $v'$ may correspond. 
The change of state is obviously completely determined if we know 1) the change in temperature 2) the displacements that a point, which initially had the right-angled coordinates $x_0$, $y_0$, $z_0$, has undergone in the directions of the coordinate axes, and which we want to denote by $\xi$, $\eta$, $\zeta$, which are  quantities (initially) to be thought of as functions of $x_0$, $y_0$, $z_0$.
Then the position of this point after the change of state is determined by the coordinates:
\vspace*{0mm} \begin{align}
 x \; = \; x_0 \:+\: \xi
 \quad \quad \quad \quad \quad \quad
 y \; = \; y_0 \:+\: \eta
 \quad \quad \quad \quad \quad \quad
 z \; = \; z_0 \:+\: \zeta
 \; .
\nonumber \end{align} 
With the help of these equations, the quantities $\xi$, $\eta$, $\zeta$ can of course also be represented as functions of the current coordinates $x$, $y$, $z$, which will be discussed later.

If a new state of equilibrium is now reached, we can immediately specify the conditions for which the changes that have occurred in the state of the body must necessarily be fulfilled in this case. 
Firstly, experience shows that the necessary condition for equilibrium is that the temperature in the whole body is the same. 
So if the temperature has changed from $T$ to $T'$ at any point, the same must be the case at all other points, which we can express by: 
   $$dT' \;=\; 0 \; ,$$ 
where the sign ``$d$\,'' denotes the transition from one point of the body to an infinitely neighbouring one.

Furthermore, a necessary condition of equilibrium is that every body element, both internally and on the surface, if it is thought of as rigid, is kept in equilibrium by the forces acting on its lateral surfaces. Let us denote, as is customary in the theory of elasticity, the components of the elastic force acting on a surface element $d\omega$ parallel to the $YZ$ plane as 
$X_x\:d\omega$, $Y_x\:d\omega$, $Z_x\:d\omega$, 
and if we define the quantities 
$X_y$, $Y_y$, $Z_y$ and $X_z$, $Y_z$, $Z_z$ analogously, 
we have the known equilibrium conditions for a solid element in the interior:
\vspace*{1mm} 
\begin{equation}
\left.
\begin{aligned}
  \frac{\partial\, X_x}{\partial\, x} 
    \; + \;
  \frac{\partial \,{\color{red}X_y}}{\partial\, y} 
    \; + \;\,
  \frac{\partial\, {\color{teal}X_z}}{\partial\, z} 
  & \: = \;0  \; ,
  \vspace*{2mm} 
\\
  \frac{\partial\, {\color{red}Y_x}}{\partial\, x} 
    \,\:\; + \;
  \frac{\partial\, Y_y}{\partial\, y} 
    \,\:\; + \;
  \frac{\partial\, {\color{blue}Y_z}}{\partial\, z} 
  & \: = \;0   \; ,
  \vspace*{2mm} 
\\
  \frac{\partial\, {\color{teal}Z_x}}{\partial\, x} 
    \,\; + \;
  \frac{\partial\, {\color{blue}Z_y}}{\partial\, y} 
    \,\; + \;
  \frac{\partial\, Z_z}{\partial\, z} 
  & \: = \;0   \; ,
  \vspace*{2mm} 
\end{aligned}
\;\;\;\;\;\;\;\;
\right\} 
\label{Eq_1} 
\end{equation}
where {\color{red}{\it(I have coloured the symmetric components / P. Marquet)}}:
\vspace*{-1mm} \begin{align}
 {\color{red}X_y \; = \; Y_x}
 \quad \quad \quad
 {\color{blue}Y_z \; = \; Z_y}
 \quad \quad \quad
 {\color{teal}Z_x \; = \; X_z}
 \; .
\label{Eq_2} \end{align} 
Here the components of the elastic forces are to be viewed as functions of $x$, $y$, $z$ alone, because if the state of the body is completely given, as we assume here, then for each point $x$, $y$, $z$ we also know the elastic forces that act on the individual surface elements placed through this point.

The following conditions apply to the points on the surface: 
\vspace*{-1mm} 
\begin{equation}
\left.
\begin{aligned}
    \Xi
    \; + \;
  \alpha \; X_x
    \; + \;
  \beta \; {\color{red}X_y}
    \; + \;
  \gamma \; {\color{teal}X_z}
  & \: = \;0   \; ,
  \vspace*{2mm} 
\\
    H
    \; + \;
  \alpha \; {\color{red}Y_x}
    \,\:\; + \;
  \beta \; Y_y
    \,\:\; + \;
  \gamma \; {\color{blue}Y_z}
  & \: = \;0   \; ,
  \vspace*{2mm} 
\\
    Z
    \; + \;
  \alpha \; {\color{teal}Z_x}
    \,\; + \;
  \beta \; {\color{blue}Z_y}
    \,\; + \;
  \gamma \; Z_z
  & \: = \;0   \; .
  \vspace*{2mm} 
\end{aligned}
\;\;\;\;\;\;\;\;
\right\} 
\label{Eq_3} 
\end{equation}
Here the components of the force acting on an element $d\sigma$ of the surface from the outside are denoted by $\Xi\:d\sigma$, $H\:d\sigma$, $Z\:d\sigma$, furthermore the direction cosines of the external normal are denoted by $\alpha$, $\beta$, $\gamma$. Then the pressure components of the internal elastic forces $X_x$, $Y_y$, $Z_z$ have positive values when pressure is exerted on the body from the outside and negative values when tension is exerted.

While up to this point the equations established for the elastic forces do not differ in any way from those developed in the theory of elasticity, if the temperature is regarded as constant for all changes of state, a very significant difference can soon be recognised between those and these expressions.

Let us imagine that the body passes from the state of equilibrium just considered to some infinitely neighbouring state of equilibrium, and then calculate the elementary work performed by the elastic forces. The latter state of equilibrium is completely determined by the change in temperature $\delta T'$ and the displacements $\delta\,\xi$, $\delta\,\eta$, $\delta\,\zeta$ in the directions of the $3$ coordinate axes, 
endured by 
a point that previously had the coordinates $x_0+\xi$, $y_0+\eta$, $z_0+\zeta$.

Since both at the beginning and at the end of the infinitely small change of state each body element is in equilibrium inside under the action of the elastic forces, the work of the elastic forces inside the body is $=0$, and therefore the total work of the elastic forces is limited to that of the elastic forces on the surface, because these do not in and of themselves keep the body elements in equilibrium.

The components of the elastic forces acting on a body element at the surface are, if $d\sigma$ denotes the area of the corresponding surface element, according to equations~(\ref{Eq_3}):
\vspace*{-1mm} 
\begin{equation}
\begin{aligned}
   \left(\:
  \alpha \; X_x
    \; + \;
  \beta \; {\color{red}X_y}
    \; + \;
  \gamma \; {\color{teal}X_z}
  \:\right) 
  \;d\sigma  \; ,
  \vspace*{2mm} 
\\
    \left(\:
  \alpha \; {\color{red}Y_x}
    \,\:\; + \;
  \beta \; Y_y
    \,\:\; + \;
  \gamma \; {\color{blue}Y_z}
  \:\right)
  \;d\sigma  \; ,
  \vspace*{2mm} 
\\
    \left(\:
  \alpha \; {\color{teal}Z_x}
    \,\; + \;
  \beta \; {\color{blue}Z_y}
    \,\; + \;
  \gamma \; Z_z
  \:\right)
  \;d\sigma  \; .
  \vspace*{2mm} 
\end{aligned}
\;\;\;\;\;\;\;\;
\nonumber
\end{equation}


Multiplying these $3$ expressions by the corresponding displacements $\delta\,\xi$, $\delta\,\eta$, $\delta\,\zeta$ of the body element along the $3$ coordinate axes, we obtain the work of the elastic forces on the surface by adding the $3$ products, and by integrating over the whole surface, finally by changing the sign: 
\vspace*{-1mm} 
\begin{equation}
\delta\,\Phi \;=\;
 - \bigintss d\sigma \;
\left\{ \;
\begin{aligned}
   \;\;\;\:\left(\:
  \alpha \; X_x
    \; + \;
  \beta \; {\color{red}X_y}
    \; + \;
  \gamma \; {\color{teal}X_z}
  \:\right) 
  \;\delta\,\xi
  \vspace*{2mm} 
\\
   +\; \left(\:
  \alpha \; {\color{red}Y_x}
    \,\:\; + \;
  \beta \; Y_y
    \,\:\; + \;
  \gamma \; {\color{blue}Y_z}
  \:\right)
  \;\delta\,\eta
  \vspace*{2mm} 
\\
   +\; \left(\:
  \alpha \; {\color{teal}Z_x}
    \,\; + \;
  \beta \; {\color{blue}Z_y}
    \,\; + \;
  \gamma \; Z_z
  \:\right)
  \;\delta\,\gamma
  \vspace*{2mm} 
\end{aligned}
\;
\right\} \; .
\label{Eq_4} 
\end{equation}
As far as the change of sign is concerned, this is due to the fact that the quantity of work is calculated in the opposite sense in the mechanical theory of heat than in mechanics. For if a point moves under the influence and in the direction of a force, then in mechanics the work is represented by the positive product of force and displacement, whereas in the mechanical theory of heat it is represented by the negative product of force and displacement, because (according to the usual way of expression) no work is done by the process just described, but work is consumed.

The expression (\ref{Eq_4}), which according to the above now represents the total work of the elastic forces of the body for an infinitely small change of state, can be transformed from a surface integral into a physical integral in a known way, for example by having for a member of this expression: 
\vspace*{-1mm} 
\begin{equation}
- \bigintssss d\sigma \:.\: 
 \left(\: \alpha \; X_x \;\delta\,\xi \:\right) 
 \; = \;
- \bigintssss\!\!\bigintssss\!\!\bigintssss 
  dx\:dy\:dz \:.\:
 \left(\: 
    \frac{\partial\, X_x}{\partial\, x} \;\delta\,\xi 
    \:+\:
    X_x \:.\: \delta\:\frac{\partial\, \xi}{\partial\, x} 
 \:\right) 
\; .
\label{Eq_5} 
\end{equation}

If you now adds the $9$ members of the expression (\ref{Eq_4}), if you have transformed them all according to (\ref{Eq_5}), then this sum simplifies in that all the terms multiplied by $\delta\,\xi$, $\delta\,\eta$, $\delta\,\zeta$ cancel each other out according to equations (\ref{Eq_1}), and then taking into account the equations (\ref{Eq_2}) we get:
\vspace*{-1mm} 
\begin{equation}
\delta\,\Phi \;=\;
- \bigintssss\!\!\bigintssss\!\!\bigintssss 
  dx\:dy\:dz \:.\:
\left\{ \;
\begin{aligned}
  X_x \:.\: \delta\:\frac{\partial\, \xi}{\partial\, x}
    \; + \; 
  {\color{red}X_y} \;.\: 
  \delta \left(\:
  \frac{\partial\, \xi}{\partial\, y}
    \; + \;   
  \frac{\partial\, \eta}{\partial\, x}
  \:\right)
  \vspace*{2mm} 
\\
  \:+\:
  Y_y \:.\: \delta\:\frac{\partial\, \eta}{\partial\, y}
    \; + \; 
  {\color{blue}Y_z} \;.\: 
  \delta \left(\:
  \frac{\partial\, \eta}{\partial z}
    \; + \;   
  \frac{\partial\, \zeta}{\partial y}
  \:\right)\;
  \vspace*{2mm} 
\\
  \:+\:
  Z_z \:.\: \delta\:\frac{\partial\, \zeta}{\partial\, z}
    \; + \; 
  {\color{teal}Z_x} \;.\: 
  \delta \left(\:
  \frac{\partial\, \zeta}{\partial x}
    \; + \;   
  \frac{\partial\, \xi}{\partial z}
  \:\right)\,
  \vspace*{2mm} 
\end{aligned}
\;
\right\} \; .
\nonumber 
\end{equation}
If we finally set 
$$ 
{\rm x}_x \;=\; \frac{\partial\, \xi}{\partial\, x}  \; , 
\quad\quad\quad
{\rm y}_y \;=\; \frac{\partial\, \eta}{\partial\, y}   \; ,
\quad\quad\quad
{\rm z}_z \;=\; \frac{\partial\, \zeta}{\partial\, z}   \; ,
$$
 and 
$$
{\rm x}_y \:=\: {\rm y}_x \:=\: 
     \frac{\partial\, \xi}{\partial\, y} 
     \:+\:
     \frac{\partial\, \eta}{\partial\, x}   \; ,
\quad\quad\quad
{\rm y}_z \:=\: {\rm z}_y \:=\: 
     \frac{\partial\, \eta}{\partial\, z} 
     \:+\:
     \frac{\partial\, \zeta}{\partial\, y}   \; ,
\quad\quad\quad
{\rm z}_x \:=\: {\rm x}_z \:=\: 
     \frac{\partial\, \zeta}{\partial\, x} 
     \:+\:
     \frac{\partial\, \xi}{\partial\, z}   \; ,
$$
as is customary, and introduce the mass element $dM$ instead of the volume element $dx\:dy\:dz$, so that 
$$dx\;dy\;dz \;=\; dM \:.\: v' \; $$ 
(where $v'$ denotes the volume of the mass unit, thought of as a function of $x$, $y$, $z$),
we have as the total work of the elastic forces: 
\vspace*{-1mm} 
\begin{equation}
\delta\,\Phi \;=\;
- \bigintssss\!\!\bigintssss\!\!\bigintssss 
  dM \:.\: v' \:.\:
\left\{ \;
\begin{aligned}
  X_x \:.\: \delta{\rm x}_x
    \; + \;\; 
  {\color{red}X_y} \;.\: \delta{\rm x}_y
  \vspace*{2mm} 
\\
  \:+\:
  Y_y \:.\: \delta{\rm y}_y
    \;\; + \;\; 
  {\color{blue}Y_z} \;.\: \delta{\rm y}_z \:
  \vspace*{2mm} 
\\
  \:+\:
  Z_z \:.\: \delta{\rm z}_z
    \;\; + \;\; 
  {\color{teal}Z_x} \;.\: \delta{\rm z}_x \:
  \vspace*{2mm} 
\end{aligned}
\;
\right\} \; .
\label{Eq_6}
\end{equation}
Although this expression is outwardly similar to that which represents the work of the elastic forces in the theory of elasticity, it differs from that in the essential point that it does not represent the complete variation of a certain function $\Phi$. Because if the components of the elastic forces are conceived in it as functions of the quantities ${\rm x}_x\,, \: {\rm y}_y \:.\:.\:.$, they do not depend on these quantities alone, but in any case also on the temperature.
But since the variation of the temperature $\delta T'$ does not appear in the expression, it cannot be the complete variation of a function. In other words: 
\vspace*{-4mm} 
\begin{center}
the elastic forces, considered as dependent on temperature, have no potential. \\
The significance of this fact will also be shown below in a different light.
\end{center}
\vspace*{-3mm}
If, as a result of the loss of potential, a loss of evidence for the theory is suffered to a certain extent, this is completely compensated for --and even appears to be essential-- as soon as the two main theorems of the mechanical heat theory are applied to the present case.

\subsection{\underline{Application of the first law} (p.8-14)}
\vspace*{-1mm}

Let us start from the state of equilibrium characterised by the temperature $T'$ and the displacements $\xi\,, \eta\,, \zeta$, and imagine it to be infinitely little changed by the fact that an infinitely small quantity of heat $\delta\,Q$ (measured mechanically) is supplied to the body from outside, while at the same time some forces act on its surface. 
In this way, a second equilibrium state is achieved, which corresponds to the temperature change $\delta\,T'$ and the displacement changes 
$\delta\,\xi\,, \delta\,\eta\,, \delta\,\zeta\,$.

Then, according to the first law, the heat $\delta\,Q$ supplied from outside is used: 1) to increase the energy $U$ of the body (heat + internal work); and 2) to perform external work; so that if we denote the latter by $\delta\,W$ we have: 
\vspace*{-3mm} \begin{align}
 \delta\,Q \; = \; \delta\,U \:+\: \delta\,W
 \; .
\label{Eq_7} \end{align} 
In this equation, $\delta\,U$ is the complete variation of a certain function of those quantities that determine the state of the body, while $\delta\,W$ and $\delta\,Q$ generally do not have this property.

Let us now try to derive the expression for $U$. This can be done directly under a condition which is also made in the theory of elasticity, namely that the value of $U$ for the initial state, as we defined it at the beginning of this section, is known, and that the displacements $\xi\,, \eta\,, \zeta\,$ which determine the equilibrium state under consideration are very small for all points, relative to the dimensions of the body.  Similarly, we assume that the temperature $T'$ differs very little from the initial temperature $T$, so that if we set: 
\vspace*{-3mm} \begin{align}
 T' \;=\; T \;+\; \tau 
 \; ,
\label{Eq_8} \end{align} 
the quantity $\tau$ is very small.

If we now compile the variables on which the energy $U$ depends, these are: 1) the temperature change $\tau$; and 2) the displacements $\xi\,, \eta\,, \zeta\,$ as functions of $x,y,z$, because these completely determine the state of the body. In addition, of course, certain constant values are included in the expression of $U$, which depend on the selected initial state.

If we first form the expression for the energy $dU$ of a mass element $dM$, it will be proportional to $dM$ and will also depend on $\tau$ and the values that the quantities $\xi\,, \eta\,, \zeta\,$ have for the different points of this element.

But since we assume, as is also done in the theory of elasticity, that the effects of the elastic forces inside the body always extend only to infinitely neighbouring points, it follows that of the quantities $\xi\,, \eta\,, \zeta\,$ at most the first differential quotients 
according to $x\,, y\,, z$ come into consideration, because the members multiplied by higher derivatives are infinitesimally small compared with the former.

Furthermore, $d\,U$ cannot depend on the absolute magnitudes of the displacements $\xi\,, \eta\,, \zeta\,$ of the element, because if the element is displaced uniformly in space at all its points, 
its energy does not change.

Therefore, the only remaining quantities determining the value of the energy for a mass element are: $\tau$ and the $9$ first derivatives of $\xi\,, \eta\,, \zeta\,$ with respect to $x\,, y\,, z$.

However, the energy does not change even if the body rotates around any rectilinear axis at a constant temperature without suffering a deformation. From this it follows, as is known from the theory of elasticity, that the nine ditferential quotients of $\xi\,, \eta\,, \zeta\,$ only occur in the following $6$ forms: 
\vspace*{0mm} \begin{align}
 {\rm x}_x &\;=\;  \frac{\partial\, \xi}{\partial\, x} \; ,
&{\rm y}_y &\;=\; \frac{\partial\, \eta}{\partial\, y}  \; ,
&{\rm z}_z &\;=\; \frac{\partial\, \zeta}{\partial\, z}  \; ,
\nonumber \\
 {\rm x}_y \:=\: {\rm y}_x 
          &\:=\: 
     \frac{\partial\, \xi}{\partial\, y} 
     \:+\:
     \frac{\partial\, \eta}{\partial\, x}  \; ,
&{\rm y}_z \:=\: {\rm z}_y 
          &\:=\: 
     \frac{\partial\, \eta}{\partial\, z} 
     \:+\:
     \frac{\partial\, \zeta}{\partial\, y}  \; ,
&{\rm z}_x \:=\: {\rm x}_z 
          &\:=\: 
     \frac{\partial\, \zeta}{\partial\, x} 
     \:+\:
     \frac{\partial\, \xi}{\partial\, z}  \; , 
\nonumber \end{align} 
in terms of {\color{red}{\it(the same variables/P. Marquet)}}
$\;{\rm x}_x\,, \: {\rm y}_y \:.\:.\:.$ 
we have already introduced above.

Therefore, with an infinitely small change of state, the energy $d\,U$ of the element $d\,M$ will change by a quantity $\delta\,d\,U$, which consists of $7$ members, each of which is multiplied by one of the $7$ variation terms  
$\delta\,\tau, \: \delta\,{\rm x}_x\,, \: \delta\,{\rm y}_y \:.\:.\:.$

The coefficients of these variations are of course generally themselves functions of the $7$ variables 
$\tau,\:{\rm x}_x\,,\:{\rm y}_y \:.\:.\:.$
However, since these are very small according to the assumption, they are linear functions of these $7$ variables, and therefore $d\,U$ is a quadratic function of them, whose expression we now want to set up.

It first contains a constant element that depends only on the chosen initial state, then elements that depend linearly, and finally elements that depend quadratically on the variables.

It is clear that, since the solid is isotropic, the function must be composed symmetrically with respect to those variables whose meaning differs only by the choice of the coordinate axes, that is, with respect to 
${\rm x}_x\,,\:{\rm y}_y\,,\:{\rm z}_z$
on the one hand, and with respect to 
${\rm x}_y\,,\:{\rm y}_z\,,\:{\rm z}_x$
on the other. 
As a result, the linear elements are reduced to two, one of which has $\tau$, and the other of which has 
$({\rm x}_x\,+\,{\rm y}_y\,+\,{\rm z}_z)$
as a factor, because the element multiplied by 
$({\rm x}_x\,+\,{\rm y}_y\,+\,{\rm z}_z)$, 
which alone would still be possible, must disappear. 
If the positive direction of the x-axis is swapped with the negative direction, 
${\rm x}_y$ changes into $-\,{\rm x}_y$ and
${\rm z}_x$ changes into $-\,{\rm z}_x$,
while ${\rm y}_z$ retains its value. 
And 
since this does not change the value of the energy, the coefficient of $({\rm x}_x\,+\,{\rm z}_z)$, and therefore also that of $({\rm x}_x\,+\,{\rm y}_y\,+\,{\rm z}_z)$, must have the value $0$.

\newpage
Finally, the quadratic terms
are reduced to 5, namely:
\vspace*{1mm} \\ 
\hspace*{15mm} $1.\;\;$one member with $\;\tau^2$\,, 
\vspace*{1mm} \\ 
\hspace*{15mm} $2.\;\;$one member with 
$\;\tau\;(\,{\rm x}_x\,+\,{\rm y}_y\,+\,{\rm z}_z\,)$\,, 
\vspace*{1mm} \\ 
\hspace*{15mm} $3.\;\;$one member with 
$\;(\,{\rm x}_x\,+\,{\rm y}_y\,+\,{\rm z}_z\,)^2$\,; 
\vspace*{1mm} \\ 
\hspace*{15mm} $4.\;\;$one member with  
$\;(\,{\rm x}_x^2\,+\,{\rm y}_y^2\,+\,{\rm z}_z^2\,)$\,; 
\vspace*{1mm} \\ 
\hspace*{15mm} $5.\;\;$one member with   
$\;(\,{\rm x}_y^2\,+\,{\rm y}_z^2\,+\,{\rm z}_x^2\,)$\,. 

The coefficient of the last member 5. is half the coefficient of the last but one 4. 
-- The considerations on which this reduction are based (together with the last condition mentioned) are founded 
on the condition that the body is isotropic, and are made in the same way as in the elasticity theory when determining the potential of the elastic forces, so that we don't need to go into it in more detail here.

If we introduce the dilation of the volume unit
$$ \Theta \;=\; \,{\rm x}_x\,+\,{\rm y}_y\,+\,{\rm z}_z\, $$
as an abbreviation, 
we thus finally obtain the following expression for the energy $d\,U$ of the mass element $d\,M$: 
\vspace*{-1mm} 
\begin{equation}
d\,U \;=\;
  d\,M \:.\: 
\left\{ \;
\begin{aligned}
  const. \:+\: k \:.\: \tau \:+\: l \:.\: \Theta 
  \hspace*{17mm} 
\\
  \:+\: \frac{m}{2} \:.\: \tau^2
  \:+\: p \:.\: \tau \:.\: \Theta 
  \:+\: \frac{q}{2} \:.\: \Theta^2 
  \hspace*{11mm}  
\\
  \:+\: r \:.\: 
  (\,{\rm x}_x^2\,+\,{\rm y}_y^2\,+\,{\rm z}_z^2\,)
  \:+\: \frac{r}{2} \:.\:
  (\,{\rm x}_y^2\,+\,{\rm y}_z^2\,+\,{\rm z}_x^2\,) 
  \hspace*{0mm}  
\end{aligned}
\;
\right\} \; .
\label{Eq_9}
\end{equation}
\vspace*{1mm} 

If we denote the expression multiplied by $d\,M$ briefly by $u$ (the specific energy), we obtain the energy of the whole body: \vspace*{-1mm} 
\begin{equation}
  U \;=\; \bigintssss d\,M \:.\: u \; .
\label{Eq_10}
\end{equation}
Here, the $6$ constants $k,l,m,p,q,r$ are determined by the initial state of the body and are therefore completely independent of the position of the mass element $d\,M$, so that they can be placed in front of the integral sign, just like $\tau$.

Now we are also able to represent the expression $\delta\,U$ occurring in equation (\ref{Eq_7}) as a function of the determinants of the state. From (\ref{Eq_10}) we have: 
$$  \delta\,U \;=\; \bigintssss d\,M \:.\: \delta\,u \; , $$ 
or: 
\vspace*{1mm} 
\begin{equation}
\delta\,U \;=\;
  \bigintssss d\,M \:.\: 
\left\{ \;
\begin{aligned}
  \frac{\partial\,u}{\partial\,\tau} \:.\: \delta\,\tau
  \:+\: 
  \frac{\partial\,u}{\partial\,{\rm x}_x}\:.\:\delta\,{\rm x}_x
  \:+\: 
  \frac{\partial\,u}{\partial\,{\rm y}_y}\:.\:\delta\,{\rm y}_y
  \:+\: 
  \frac{\partial\,u}{\partial\,{\rm z}_z}\:.\:\delta\,{\rm z}_z
  \hspace*{0mm}  
\\
  \:+\: 
  \frac{\partial\,u}{\partial\,{\rm x}_y}\:.\:\delta\,{\rm x}_y
  \:+\: 
  \frac{\partial\,u}{\partial\,{\rm y}_z}\:.\:\delta\,{\rm y}_z
  \:+\: 
  \frac{\partial\,u}{\partial\,{\rm z}_x}\:.\:\delta\,{\rm z}_x
  \hspace*{0mm}  
\end{aligned}
\;
\right\} \; .
\label{Eq_11}
\end{equation}
\vspace*{1mm} 

Here, as can be seen from (\ref{Eq_9}):
\vspace*{-3mm} 
\begin{equation}
\left. \;
\begin{aligned}
  \frac{\partial\,u}{\partial\,\tau} \; = \;
  k \:+\: m \:.\: \tau \:+\: p \:.\: \Theta
  \hspace*{0mm}  
\\
  \frac{\partial\,u}{\partial\,{\rm x}_x} \; = \;
  l \:+\: p \:.\: \tau \:+\: q \:.\: \Theta 
    \:+\: 2\:.\: r \:.\: {\rm x}_x
  \hspace*{0mm}  
\\
  \frac{\partial\,u}{\partial\,{\rm y}_y} \; = \;
  l \:+\: p \:.\: \tau \:+\: q \:.\: \Theta 
    \:+\: 2\:.\: r \:.\: {\rm y}_y
  \hspace*{0mm}  
\\
  \frac{\partial\,u}{\partial\,{\rm z}_z} \; = \;
  l \:+\: p \:.\: \tau \:+\: q \:.\: \Theta 
    \:+\: 2\:.\: r \:.\: {\rm z}_z
  \hspace*{0mm}  
\\
  \frac{\partial\,u}{\partial\,{\rm x}_y} \; = \;
  r \:.\: {\rm x}_y
  \quad\quad
  \frac{\partial\,u}{\partial\,{\rm y}_z} \; = \;
  r \:.\: {\rm y}_z
  \quad\quad
  \frac{\partial\,u}{\partial\,{\rm z}_x} \; = \;
  r \:.\: {\rm z}_x
  \hspace*{0mm}  
\end{aligned}
\;
\right\} \; .
\label{Eq_12}
\end{equation}


The first law of mechanical heat theory offers no clues for drawing direct conclusions from the value of the energy $U$ to the magnitude of the elastic forces themselves. Rather, this requires the second law, the application of which is described below.

The second element on the right-hand side of equation (\ref{Eq_7}), which represents the external work done $\delta\,W$, can, in contrast to the first element $\delta\,U$, have very different values depending on the external circumstances under which the infinitesimal change of state takes place. However, we now want to calculate $\delta\,W$ for the special case that the external work performed consists of overcoming those external forces that keep the body in equilibrium. For this purpose, the principle of virtual velocities can be applied with advantage: since in this case the acting external forces together with the elastic forces keep the system in equilibrium, the total work done by all acting forces is $=0$, i.e. 
$$\delta\,W \:+\: \delta\,\Phi \;=\;  0  \; , $$
where $\delta\,\Phi$, the work of the elastic forces, is given by equation (\ref{Eq_6}). Consequently:
\vspace*{-1mm} 
$$\delta\,W \;=\; - \: \delta\,\Phi \; .$$

If this value of $\delta\,W$ is substituted in equation (\ref{Eq_7}), the result is: 
\vspace*{-1mm} \begin{align}
  \delta\,Q \;=\; \delta\,U - \: \delta\,\Phi \; .
\label{Eq_13} \end{align} 
In this equation, $\delta\,Q$ is expressed by two quantities which are completely determined by the state of the body and the infinitesimal change of state. However, since, as we have seen above, $\delta\,\Phi$ does not, like $\delta\,U$, have the property of being the complete variation of a certain function, $\delta\,Q$ does not represent such a variation either, i.e. the equation is not generally integrable in this form, and thus $\delta\,Q$ cannot be integrable according to the principles of mechanical heat theory.
As is well known, the amount of heat that must be supplied to a body from outside so that it can move from one state to another state depends not only on the two states, but also on the way in which it moves from one state to the other brought to another state. 
From this we recognise the fact that $\delta\,\Phi$ is not the complete variation of a function, as a necessary consequence of the theory. 
The second law now teaches us the factor by which equation (\ref{Eq_13}) must be multiplied in order to be generally integrable, and this is, as we know, the reciprocal value of the absolute temperature $(1/T)$.

\subsection{\underline{Application of the second law} (p.14-32)}
\vspace*{-1mm}

The theorem first stated in this form by Clausius is known to be: if a body undergoes a reversible circular{\color{red}\,-{\it(cyclic)}} process in which it partly receives heat from outside or emits heat to the outside, partly is deformed by external forces, then the equation applies: 
$$ \int \frac{\delta\,Q}{T'} \;=\; 0 \; , $$ 
where $\delta\,Q$ denotes a communicated heat element, $T'$ the temperature of the same, and the integral is to be extended over the entire cycle process.
However, since this is reversible, the body has the same temperature at every moment as the amount of heat just absorbed from outside, consequently $T'$ also denotes the temperature of the body when absorbing the heat $\delta\,Q$. However, equation (\ref{Eq_13}) applies to $\delta\,Q$ because, due to the reversibility of the process, the external work consists of overcoming the external forces that are currently keeping the body in equilibrium. Therefore we have:
$$ \int \frac{\delta\,U \:-\: \delta\,\Phi}{T'} \;=\; 0 \; . $$
This equation is composed only of determinants of the state of the body and is always valid if the two limits of the integral correspond to the same state, i.e. are identical. It follows directly from this that, if the limits are different, the integral has a very definite value which depends only on the limits, but not on the way in which it is integrated, i.e. 
$({\delta\,U \:-\: \delta\,\Phi})/{T'}$ 
is the complete variation of a certain function, the entropy. If we denote this by $S$, we obtain: 
\vspace*{0mm} \begin{align}
  \delta\,S \;=\; \frac{\delta\,U - \: \delta\,\Phi}{T'} \; .
\label{Eq_14} \end{align} 
Since we have already derived the expressions for 
$\delta\,U$ and $\delta\,\Phi$, 
we are immediately able to specify the conditions that this equation imposes on the two functions.

Let us first establish the form of the expression that $S$ must have. If we first form the entropy $d\,S$ of the mass element $d\,M$, we can use the same considerations as above in determining the value of $d\,U$. This is because no further assumptions were made there other than that the body is isotropic and that the temperature change $\tau$ and the displacements $\xi\,, \eta\,, \zeta\,$ calculated from the initial state are very small. We can therefore also write the analogue of the equation (\ref{Eq_9}): 
\vspace*{-1mm} 
\begin{equation}
d\,S \;=\;
  d\,M \:.\: 
\left\{ \;
\begin{aligned}
  const. \:+\: k' \:.\: \tau \:+\: l' \:.\: \Theta 
  \hspace*{17mm} 
\\
  \:+\: \frac{m'}{2} \:.\: \tau^2
  \:+\: p' \:.\: \tau \:.\: \Theta 
  \:+\: \frac{q'}{2} \:.\: \Theta^2 
  \hspace*{11mm}  
\\
  \:+\: r' \:.\: 
  (\,{\rm x}_x^2\,+\,{\rm y}_y^2\,+\,{\rm z}_z^2\,)
  \:+\: \frac{r'}{2} \:.\:
  (\,{\rm x}_y^2\,+\,{\rm y}_z^2\,+\,{\rm z}_x^2\,) 
  \hspace*{0mm}  
\end{aligned}
\;
\right\} \; .
\label{Eq_15}
\end{equation}
\vspace*{1mm}
The constants $k',\,l',\,m',\,p',\,q',\,r'$ are characterised by the initial state and therefore have the same values for all points of the body.

%
If we use the designation $s$ (specific entropy) 
for the expression multiplied by $d\,M$, 
the entropy of the body is
\vspace*{-3mm} 
\begin{equation}
  S \;=\; \bigintssss d\,M \:.\: s \; ,
\nonumber 
\end{equation}
and from this follows the variation of the entropy for an infinitely small change of state 
\vspace*{1mm} 
\begin{equation}
\delta\,S 
 \;=\;
 \bigintssss d\,M \:.\: \delta\,s 
 \;=\;
  \bigintssss d\,M \:.\: 
\left\{ \;
\begin{aligned}
  \frac{\partial\,s}{\partial\,\tau} \:.\: \delta\,\tau
  \:+\: 
  \frac{\partial\,s}{\partial\,{\rm x}_x}\:.\:\delta\,{\rm x}_x
  \:+\: 
  \frac{\partial\,s}{\partial\,{\rm y}_y}\:.\:\delta\,{\rm y}_y
  \:+\: 
  \frac{\partial\,s}{\partial\,{\rm z}_z}\:.\:\delta\,{\rm z}_z
  \hspace*{0mm}  
\\
  \:+\: 
  \frac{\partial\,s}{\partial\,{\rm x}_y}\:.\:\delta\,{\rm x}_y
  \:+\: 
  \frac{\partial\,s}{\partial\,{\rm y}_z}\:.\:\delta\,{\rm y}_z
  \:+\: 
  \frac{\partial\,s}{\partial\,{\rm z}_x}\:.\:\delta\,{\rm z}_x
  \hspace*{0mm}  
\end{aligned}
\;
\right\} \; ,
\label{Eq_16}
\end{equation}
whereby the values of the derivatives of $s$ result directly from the right-hand side of equation (\ref{Eq_15}) and differ from the derivatives of $U$, see (\ref{Eq_12}), only by the (accentuated) constants.

If we now consider the following equation from (14): 
\vspace*{-1mm} \begin{align}
  \delta\,U - \: \delta\,\Phi \;=\; {T'} \:.\: \delta\,S \; 
\nonumber 
\end{align} 
and imagine their values substituted for 
$\delta\,U$, $\delta\,\Phi$ and $\delta\,S$ 
from (\ref{Eq_11}), (\ref{Eq_6}) and (\ref{Eq_16}), then each side of the equation consists of an integral to be extended over all mass elements of the body, since on the right-hand side $T'$ may be placed after the integral sign. However, since the change of state is completely arbitrary and the displacements of the individual mass elements do not depend on each other, the equation is also valid for each mass element taken separately. Therefore, by omitting the integral sign and the common factor $\delta\,M$, we obtain the following relation: 
\vspace*{1mm} 
\begin{equation}
\begin{aligned}
  \frac{\partial\,u}{\partial\,\tau} \:.\: \delta\,\tau
  \:+\: 
  \frac{\partial\,u}{\partial\,{\rm x}_x}\:.\:\delta\,{\rm x}_x
  \:+\: 
  \frac{\partial\,u}{\partial\,{\rm y}_y}\:.\:\delta\,{\rm y}_y
  \:+\: 
  \frac{\partial\,u}{\partial\,{\rm z}_z}\:.\:\delta\,{\rm z}_z
  \hspace*{0mm}  
\\
  \:+\: 
  \frac{\partial\,u}{\partial\,{\rm x}_y}\:.\:\delta\,{\rm x}_y
  \:+\: 
  \frac{\partial\,u}{\partial\,{\rm y}_z}\:.\:\delta\,{\rm y}_z
  \:+\: 
  \frac{\partial\,u}{\partial\,{\rm z}_x}\:.\:\delta\,{\rm z}_x
  \hspace*{0mm}  
\\
  \;+\; 
  v'\:.\:X_x\:.\:\delta\,{\rm x}_x
  \;+\; 
  v'\:.\:Y_y\:.\:\delta\,{\rm y}_y
  \;+\;
  v'\:.\:Z_z\:.\:\delta\,{\rm z}_z
\\
  \;+\; 
  v'\:.\:X_y\:.\:\delta\,{\rm x}_y
  \;+\; 
  v'\:.\:Y_z\:.\:\delta\,{\rm y}_z
  \;+\;
  v'\:.\:Z_x\:.\:\delta\,{\rm z}_x
\\
  \;=\; T' \:.\:
  \frac{\partial\,s}{\partial\,\tau} \:.\: \delta\,\tau
  \:+\: T' \:.\:
  \frac{\partial\,s}{\partial\,{\rm x}_x}\:.\:\delta\,{\rm x}_x
  \:+\: T' \:.\:
  \frac{\partial\,s}{\partial\,{\rm y}_y}\:.\:\delta\,{\rm y}_y
  \:+\: T' \:.\:
  \frac{\partial\,s}{\partial\,{\rm z}_z}\:.\:\delta\,{\rm z}_z
  \hspace*{0mm}  
\\
  \:+\:  T' \:.\:
  \frac{\partial\,s}{\partial\,{\rm x}_y}\:.\:\delta\,{\rm x}_y
  \:+\:  T' \:.\:
  \frac{\partial\,s}{\partial\,{\rm y}_z}\:.\:\delta\,{\rm y}_z
  \:+\:  T' \:.\:
  \frac{\partial\,s}{\partial\,{\rm z}_x}\:.\:\delta\,{\rm z}_x
  \hspace*{0mm} 
  \; . 
\end{aligned}
\nonumber 
\end{equation}
Since the individual variations are independent of each other, this results in the following $7$ equations:
\vspace*{1mm} 
\begin{align}
  \frac{\partial\,u}{\partial\,\tau} 
  \hspace*{19mm} 
  & \:=\; T' \:.\:
  \frac{\partial\,s}{\partial\,\tau} 
  \; , \label{Eq_17} \\
  \frac{\partial\,u}{\partial\,{\rm x}_x} 
  \;+\; 
  v'\:.\:X_x
  \hspace*{0mm} 
  & \:=\; T' \:.\:
  \frac{\partial\,s}{\partial\,{\rm x}_x} 
  \; , \label{Eq_18} \\
  \frac{\partial\,u}{\partial\,{\rm y}_y} 
  \;+\; 
  v'\:.\:Y_y
  \hspace*{0mm} 
  & \:=\; T' \:.\:
  \frac{\partial\,s}{\partial\,{\rm y}_y} 
  \; , \label{Eq_19} \\
  \frac{\partial\,u}{\partial\,{\rm z}_z} 
  \;+\; 
  v'\:.\:Z_z
  \hspace*{0mm} 
  & \:=\; T' \:.\:
  \frac{\partial\,s}{\partial\,{\rm z}_z} 
  \; , \label{Eq_20} \\
  \frac{\partial\,u}{\partial\,{\rm x}_y} 
  \;+\; 
  v'\:.\:X_y
  \hspace*{0mm} 
  & \:=\; T' \:.\:
  \frac{\partial\,s}{\partial\,{\rm x}_y} 
  \; , \label{Eq_21} \\
  \frac{\partial\,u}{\partial\,{\rm y}_z} 
  \;+\; 
  v'\:.\:Y_z
  \hspace*{0mm} 
  & \:=\; T' \:.\:
  \frac{\partial\,s}{\partial\,{\rm y}_z} 
  \; , \label{Eq_22} \\
  \frac{\partial\,u}{\partial\,{\rm z}_x} 
  \;+\; 
  v'\:.\:Z_x
  \hspace*{0mm} 
  & \:=\; T' \:.\:
  \frac{\partial\,s}{\partial\,{\rm z}_x} 
  \; . \label{Eq_23}
\end{align}

Now we substitute into these equations the values of the derivatives of $u$ and $s$, as they result directly from equations (\ref{Eq_12}) for $u$, and, with accentuation of the constants, according to (\ref{Eq_15}), also for $s$, further replace $T'$ according to equation (\ref{Eq_8}) by $T+\tau$, and finally $v'$, the volume of the mass unit, by $v\:.\:(1+\Theta)$, where $v$ denotes the specific volume in the initial state. Then the equation (\ref{Eq_17}) transforms (neglecting the small second-order quantities) into: 
$$ k \;+\; m\:.\:\tau \;+\; p\:.\:\Theta \;=\; T\:.\:k' 
  \;+\; T\:.\:m'\:.\:\tau \;+\; T\:.\:p'\:.\:\Theta 
  \;+\; k'\:.\:\tau \; .$$
However, since $\tau$ and $\Theta$ are independent of each other, the result is: 
\vspace*{1mm} 
\begin{align}
  k & \:=\; T \:.\:k'
  \; , \label{Eq_24} \\
  m & \:=\; T \:.\:m' \;+\; k'
  \; , \label{Eq_25} \\
  p & \:=\; T \:.\:p'
  \; . \label{Eq_26}
\end{align}
Moreover, from equation (\ref{Eq_18}) it follows, neglecting the 2nd-order quantities:
\vspace*{1mm} 
\begin{align}
  & l \;+\; p\:.\:\tau \;+\; q\:.\:\Theta 
    \;+\; 2\:.\:r\:.\:{\rm x}_x 
    \;+\; v\:.\:(1\:+\:\Theta)\:.\:X_x
  \; \nonumber  \\ 
  & \;=\; 
          T\:.\:l' \;+\; T\:.\:p'\:.\:\tau 
    \;+\; T\:.\:q'\:.\:\Theta 
    \;+\; 2\:.\:T\:.\:r'\:.\:{\rm x}_x
    \;+\; l'\:.\:\tau
  \; . \nonumber 
\end{align}

From this, according to (\ref{Eq_26}): 
\begin{align}
  X_x & \:=\;
        \frac{T\:.\:l' \,-\, l}{v}
  \;+\; \frac{l'}{v}\:.\:\tau
  \;-\; \frac{(q\,-\,l)\,-\,T\:.\:(q'\,-\,l')}{v}\:.\:\Theta 
  \;-\; \frac{2\:.\:(r\,-\,T\:.\:r')}{v}\:.\:{\rm x}_x 
  \; , \label{Eq_27} 
\end{align}
with the values of $Y_y$ and $Z_z$ obtained analogously from (\ref{Eq_19}) and (\ref{Eq_20}).

Finally, from (\ref{Eq_21}) follows:
$$    r\:.\:{\rm x}_y  
\;+\; v\:.\:(1\:+\:\Theta)\:.\:X_y
\;=\; T\:.\:r'\:.\:{\rm x}_y \; , $$
and from this: 
\begin{align}
  X_y & \:=\;
   -\; \frac{r\,-\,T\:.\:r'}{v}\:.\:{\rm x}_y 
  \; , \label{Eq_28} 
\end{align}
with the values of $Y_z$ and $Z_x$ obtained analogously from (\ref{Eq_22}) and (\ref{Eq_23}).

\vspace*{-3mm}
\begin{center}
--------------------------------------------------- 
\end{center}
\vspace*{-2mm}
We want to replace the constants occurring in these expressions with others, some of which are simplifications and some of which are directly accessible to experimental determination. If $\tau$ and all displacements $=0$ are set, we obtain the equations corresponding to the initial state, namely according to (\ref{Eq_27}): 
\begin{align}
  X_x & \:=\; \frac{T\:.\:l'\,-\,l}{v}
        \;=\; Y_y  \;=\; Z_z
  \; , \label{Eq_29} 
\end{align}
and according to (\ref{Eq_28}):
\vspace*{-3mm}
\begin{align}
  X_y & \:=\; 0  \;=\;  Y_z  \;=\;  Z_x  
  \; . \nonumber 
\end{align}

Thus, in the initial state, the elastic forces are reduced to a pressure that is the same everywhere and acts vertically on each surface element. If we denote this by $P$, the result is:
\vspace*{0mm}
\begin{align}
  P & \:=\; \frac{T\:.\:l'\,-\,l}{v}
  \; . \label{Eq_30} 
\end{align}
It follows from the equilibrium conditions (\ref{Eq_3}) that this initial pressure is numerically equal and opposite to the external pressure. As already emphasised there, a positive value of $P$ means a state in which the body strives to expand, whereas a negative value of $P$ means a state in which the internal elastic forces try to bring about a 
contraction. The sense in which the elastic forces are counted here is therefore the opposite of that usually assumed in the theory of elasticity.

If we now change the state of the body so that its temperature remains constant, i.e. $\tau=0$, we obtain the conditions that apply to the theory of elasticity. We will therefore retain the terms used there here by denoting in equation (\ref{Eq_27}) the constants multiplied by $\Theta$ and ${\rm x}_x$ by $\lambda$ and $2\:\mu$, respectively, so that we have: 
\vspace*{-3mm}
\begin{align}
  \lambda & \:=\; 
  \frac{(q\,-\,l)\,-\,T\:.\:(q'\,-\,l')}{v}
  \; , \nonumber \\ 
  \mu & \:=\; 
  \frac{r\,-\,T\:.\:r'}{v}
  \; . \nonumber 
\end{align}
Then (\ref{Eq_27}) with respect to (\ref{Eq_30}) becomes: 
\vspace*{-1mm}
\begin{align}
  X_x & \:=\; P 
   \;+\; \frac{l'}{v}\:.\:\tau
   \;-\; \lambda\:.\:\Theta
   \;-\; 2\:.\:\mu\:.\:{\rm x}_x 
  \; , \label{Eq_31} 
\end{align}
and (\ref{Eq_28}) becomes: 
\vspace*{-3mm}
\begin{align}
  X_y & \:=\;  
   -\; \mu\:.\:{\rm x}_y 
  \; . \label{Eq_32} 
\end{align}
The negative signs of the members multiplied by $\lambda$ and $\mu$ are determined 
by the sense of the elastic forces.

Finally, let us introduce the coefficient of thermal expansion of the body at constant pressure $P$ into the calculation. If we denote the ratio of the change in specific volume to the change in temperature at constant pressure $P$ by $\alpha$, we can write:
\vspace*{-1mm}
\begin{align}
  \alpha & \:=\; \left(\frac{v'\:-\:v}{T'\:-\:T}\right)_P
           \;=\; v\:.\:\left(\frac{\Theta}{\tau}\right)_P
  \; . \label{Eq_33} 
\end{align}
The quotient $(\Theta/\tau)_P$ is the cubic coefficient of thermal expansion. However, it is recommended for later use to introduce the quantity labelled $\alpha$ rather than this.

The value of $(\Theta/\tau)_P$ results from equation (\ref{Eq_31}) if you set $X_x \,=\, P$. Therefore: 
\vspace*{-1mm}
\begin{align}
  0 & \:=\; 
         \frac{l'}{v}\:.\:\tau
   \;-\; \lambda\:.\:\Theta
   \;-\; 2\:.\:\mu\:.\:{\rm x}_x 
  \; . \nonumber 
\end{align}
But in the case of expansion by heating at constant pressure, since all parts of the body expand proportionally, there is always: 
\vspace*{-1mm}
\begin{align}
 {\rm x}_x & \:=\; {\rm y}_y \;=\; {\rm z}_z \; , \;\;\;
  \mbox{and consequently}  \;\;\;\;
 {\rm x}_x \;=\; {\rm y}_y \;=\; {\rm z}_z \;=\; \frac{\Theta}{3} 
 \; , \nonumber  
\end{align}
and hence after substitution: 
\vspace*{-3mm}
\begin{align}
  0 & \:=\; 
         \frac{l'}{v}\:.\:\tau
   \;-\; \lambda\:.\:\Theta
   \;-\;  \frac{2}{3}\:.\:\mu\:.\:\Theta
  \; . \nonumber 
\end{align}
Therefore: 
\vspace*{-1mm}
\begin{align}
  \left(\frac{\Theta}{\tau}\right)_P
  & \:=\; \frac{3\:l'}{v\:.\:\left(3\:\lambda\:+\:2\:\mu\right)}
  \; , \nonumber 
\end{align}
and thus from (\ref{Eq_33}): 
\vspace*{-3mm}
\begin{align}
  \alpha
  & \:=\; \frac{3\:l'}{3\:\lambda\:+\:2\:\mu}
  \; , \nonumber 
\end{align}
hence:
\vspace*{-3mm}
\begin{align}
  l' & \:=\; \alpha \:.\:
    \left( \lambda\:+\:\frac{2}{3}\;\mu \right)
  \; . \label{Eq_34} 
\end{align}
By means of this value of $l'$ we obtain from (\ref{Eq_30}) the value of $l$: 
\vspace*{-1mm}
\begin{align}
  l & \:=\; 
    \alpha \;.\,
    \left(\lambda\:+\:\frac{2}{3}\;\mu \right)
    \,.\; T
  \;-\; P \:.\: v
  \; , \label{Eq_35} 
\end{align}
and from (\ref{Eq_31}) the value of $X_x$: 
\vspace*{-1mm}
\begin{align}
  X_x & \:=\; P 
   \;+\; \frac{\alpha \;.\,
    \left(3\:\lambda\:+\:2\:\mu \right)}
    {3\:v}
   \:.\:\tau
   \;-\; \lambda\:.\:\Theta
   \;-\; 2\:.\:\mu\:.\:{\rm x}_x 
  \; , \label{Eq_36} 
\end{align}
with similar values for $Y_y$ and $Z_z$.

We now also want to calculate the energy and entropy of the body by means of the constants introduced. 
For the former we have from (\ref{Eq_10}): 
\vspace*{-3mm}
\begin{equation}
  U \;=\; \bigintssss d\,M \:.\: u 
  \; . \nonumber 
\end{equation}
If we subtract the energy from the initial state
{\color{red}\it(say $u_0$)}$\,$\footnote{\color{red}\label{footnote_u0}$\:${\it I have added (here and afterwards) the reference energy term {\color{red}$u_0$}, not written nor explicitly set to $0$ by Planck (P. Marquet)}.}, 
and neglecting the small 2nd-order elements, we obtain from (\ref{Eq_9}) the  specific {\color{red}{\it(internal)}} energy
value:$\,$\footnote{\label{label_footnote_du}$\:${\color{red}\it I have written in red the alternative formulation for $\Delta u=u-u_0$ obtained from (\ref{Eq_8}) and with $\tau=(T'-T)=\Delta T$, and similarly with $\Theta=(v'-v)/V=\Delta v/v$ (P. Marquet)}.} 
\vspace*{-1mm}
\begin{equation}
  u {\color{red}\;-\; u_0} 
  \;\approx\;  k \:.\: \tau \;+\; l \:.\: \Theta
  \;\;{\color{red}\;\;\;
  \mbox{or equivalently:}\;\;\;\;  \Delta u \;\approx\; 
  k \;\: \Delta T \;+\; \frac{l}{v} \;\: \Delta v
  }
  \; . \nonumber 
\end{equation}
Substituting the value of $l$ from (\ref{Eq_35}) 
gives:$\,$\footnote{\color{red}$\:${\it The alternative formulation (in red) is obtained like in the footnote~\ref{label_footnote_du}
with $\tau=\Delta T$, and similarly with $\Theta=\Delta v/v$.  
Therefore, the alternative formulation (in red) of (\ref{Eq_37}) derived by Planck looks like the modern formulation for the (internal) energy, with thus $k$ corresponding to the ``\,specific heat at constant volume'' $c_v$ and the second term 
corresponding to ``\,$-P\:\Delta v$'' but with a modified formulation for the pressure term including a term depending on the variable $T/v$ (P. Marquet)}.} 
\vspace*{-1mm}
\begin{align}
  u {\color{red}\;-\; u_0} 
  &\;\approx\; k \:.\: \tau 
  \;+\; 
  \left\{\:
  \alpha \:.
    \left( \lambda\:+\:\frac{2}{3}\;\mu \right)
    .\: T
    \;-\; P \:.\: v
  \:\right\}
  \:.\: \Theta 
  \;  \label{Eq_37}
 \\
  {\color{red}
  \mbox{or equivalently:}\;\;\;\;
  \Delta u \;
  }
  &  
  {\color{red}
  \;\approx\; k \; \Delta T 
  \;-\; 
  \left\{\:
     P 
    \;-\; 
    \alpha \:.
    \left( \lambda\:+\:\frac{2}{3}\;\mu \right)
    .\: \frac{T}{v}
  \:\right\}
  \; \Delta v
  }
  \; . \nonumber
\end{align}
The constant $k$ cannot be determined from the observation of the elastic forces because it does not appear in their equations.

Similarly, for the specific entropy and if it is also counted from the initial state {\color{red}\it(say $s_0$)}, we obtain from 
(\ref{Eq_15})$\,$\footnote{\color{red}\label{footnote_s0}$\:${\it Similarly as for the energy, I have added (here and afterwards) the reference entropy term {\color{red}$s_0$}, not written by Planck.
I have also written in red the alternative formulation obtained from (\ref{Eq_8}) and with both $\tau/T=(T'-T)/T=\Delta T/T=\Delta \ln(T)$ and similarly $\Theta=(v'-v)/v=\Delta v/v=\Delta\ln(v)$. (P. Marquet)}.}:
\vspace*{-2mm}
\begin{equation}
  s {\color{red}\;-\; s_0} 
  \;=\;  k' \:.\: \tau \;+\; l' \:.\: \Theta
 \;\;{\color{red}\;\;\;
  \mbox{or equivalently:}\;\;\;\;  \Delta s \;\approx\; 
  k \;\: \Delta \ln(T) \;+\; l' \;\: \Delta \ln(v)
 }
  \; , \nonumber 
\end{equation}
and if we substitute their values for $k'$ and $l'$ from (\ref{Eq_24}) and 
(\ref{Eq_34}):$\,$\footnote{\color{red}\label{footnote_s0_bis}$\:${\it I have written in red the alternative formulation obtained from (\ref{Eq_8}) with $\tau/T=(T'-T)/T=\Delta T/T=\Delta \ln(T)$ and similarly with $\Theta=(v'-v)/v=\Delta v/v=\Delta \ln(v)$.
The formulation (\ref{Eq_38}) of Planck then looks like the modern formulation for the entropy, with $k$ corresponding to the ``\,specific heat at constant volume'' $c_v$, here noted $k$, and with the bracketed term corresponding to the ``\,gaz constant\,'' like in the case of perfect gas (P. Marquet)}.} 
\vspace*{-1mm}
\begin{equation}
  s {\color{red}\;-\; s_0} 
  \;\approx\; 
  \frac{k}{T} \:.\: \tau 
  \;+\; 
  \alpha \:.\:
    \left( \lambda\:+\:\frac{2}{3}\;\mu \right)
  \:.\: \Theta
  {\color{red}\;\;\;\;
  \mbox{or:}\; \Delta s \;\approx\; 
   k\;\: \Delta\ln(T)
  \;+\; 
  \left[\:
  \alpha \:.\:
    \left( \lambda\:+\:\frac{2}{3}\;\mu \right)
  \:\right]
  \: \Delta\ln(v) 
  }
  \; . \label{Eq_38}
\end{equation}
The specific heats of the body at constant volume and at constant pressure can be calculated by means of the values of $U$ and $S$. As far as the former quantity is concerned, it results directly if one takes into account that the heat input at constant volume is caused by the fact that no external work is performed, i.e. the entire amount of heat input $Q$ contributes to the increase in 
energy,$\,$\footnote{$\:${\color{red}\it I have added (here and afterwards) the (global) impact ``\,{\color{red}$U_0 = u_0 \:M$}'' of the specific reference term $u_0$, not written by Planck (P. Marquet)}.} 
i.e. 
$$Q \;=\; U {\color{red}\;-\; U_0} \; .$$
Furthermore, since for this case $\Theta=0$, the specific energy is reduced according to (\ref{Eq_37}) 
to: 
$$ u {\color{red}\;-\; u_0} \;=\; k\:.\:\tau $$ 
therefore the energy {\it(of the mass $M$)} is: 
$$ U \;=\; k\:.\:\tau\:.\:M {\color{red}\;+\; U_0} \; .  $$
We thus have for the specific heat at constant volume, if we calculate the ratio of the amount of heat added to the unit mass to the increase in temperature: 
$$ \frac{Q}{M\:.\:\tau} 
   \;=\; 
   \frac{\,U{\color{red}\;-\;U_0}\,}{\tau\:.\:M} 
   \;=\; k \; .$$
We can therefore define the constant $k$ as the specific heat at constant volume. Like all the constants used here, it is generally dependent on the selected initial state.

The value of the specific heat of the body at the constant pressure $P$ is found in a similar way. If again $Q$ denotes the heat supplied from outside, the quantity sought is most conveniently obtained from the 
equation$\,$\footnote{\color{red}\label{footnote_S0}$\:${\it Similarly as for the energy term {\color{red}$U_0$}, I have added (here and afterwards) the (global) impact ``\,{\color{red}$S_0 = s_0 \:M$}'' of the specific reference term {\color{red}$s_0$}, not written by Planck not written by Planck (P. Marquet)}.}:
\vspace*{-3mm}
$$ S {\color{red}\;-\; S_0} \;=\; \frac{Q}{T} \; , $$ 
which is valid here because the change of state under consideration is assumed to be very small, and because the external work consists in overcoming those forces which keep the body in equilibrium. We therefore have: 
\vspace*{-2mm}
\begin{equation}
  Q \;=\; T \:.\: \left( \: S {\color{red}\;-\; S_0} \: \right)
  \;=\; T \:. \bigintssss d\,M \:.\:  
  (\, s {\color{red}\:-\: s_0} \,)
  \; , \nonumber 
\end{equation}
or by means of (\ref{Eq_38}):
\vspace*{-1mm}
\begin{equation}
  Q \;=\; k \:.\: \tau \:.\: M
  \;+\; 
  \alpha \:.\:
    \left( \lambda\:+\:\frac{2}{3}\:\mu \right)
  \:.\: T \:. 
  \bigintssss d\,M \:.\:   \Theta
  \; . \nonumber 
\end{equation}
The value of $\Theta$ results from (\ref{Eq_33}) (expansion due to heating at constant pressure $P$): 
$$\Theta \;=\; \frac{\alpha \:.\: \tau}{v} \; ,$$
and hence, after substitution {\it(in the previous relationship for $Q$)}:
$$ 
  Q \;=\;
  k \:.\: \tau \:.\: M
  \;+\; 
  \frac{\alpha^2 \:.\: T}{v}
  \:.\:
    \left( \lambda\:+\:\frac{2}{3}\:\mu \right)
  \:.\: \tau \: M \; .
$$

From this, finally, the specific heat $c$ at constant pressure, i.e. the ratio of the heat supplied to the unit mass to the temperature 
increase:$\,$\footnote{\color{red}\label{footnote_c_k_T}$\:${\it Namely, a linear (affine) function of the absolute temperature $T$, and thus differently to the constant value derived in the Thesis dissertation of Planck (1879) for the specific heat (P. Marquet)}.}
$$ c \;=\; \frac{Q}{M\:.\:\tau}
  \;=\; 
  k 
  \;+\; 
 \left[\;
  \frac{\alpha^2}{v}
  \:.\:
    \left( \lambda\:+\:\frac{2}{3}\:\mu \right)
  \:\right]
  \; T
   \; ,
$$
Finally, the difference between the two specific heats is: 
\begin{equation}
  c \:-\: k 
  \;=\; 
  \frac{\alpha^2}{v}
  \:.\:
    \left( \lambda\:+\:\frac{2}{3}\:\mu \right)
  \; T
  \; . \label{Eq_39}
\end{equation}

Let us now make some brief applications of the derived equations.

\subsubsection{\underline{Solid bodies} (p.23-27)}
\vspace*{-1mm}

As initial state we can choose the natural state of the body at any mean temperature, then set $P=0$, and the expressions for the elastic forces become from (\ref{Eq_36}): 
\vspace*{-1mm}
\begin{align}
  X_x & \:=\; 
    \frac{\alpha}{v} \;.\,
    \left(\lambda\:+\:\frac{2}{3}\:\mu \right)
   \:.\:\tau
   \;-\; \lambda\:.\:\Theta
   \;-\; 2\:.\:\mu\:.\:{\rm x}_x 
  \; , \label{Eq_40} 
\end{align}
with similar values for $Y_y$ and $Z_z$.

Furthermore, from (\ref{Eq_32}): 
\vspace*{-3mm}
\begin{align}
  X_y & \:=\; -\:\mu\:.\:{\rm x}_y 
  \; , \nonumber  
\end{align}
with similar values for $Y_z$ and $Z_x$. 

The laws of cubic compression of elasticity are now to be derived under the assumption that the temperature generally changes as well.

For the cubic compression we obtain, since all parts of the body are compressed proportionally:
\vspace*{-1mm}
\begin{align}
 {\rm x}_x \;=\; {\rm y}_y \;=\; {\rm z}_z \;=\; \frac{\Theta}{3} 
  \;,\;\;\; \mbox{and}  \;\;\;\;
 {\rm x}_y \;=\; {\rm y}_z \;=\; {\rm z}_x \;=\; 0
 \; . \nonumber  
\end{align}

From this, the resulting elastic forces according to (\ref{Eq_40}) is: 
\vspace*{0mm}
\begin{align}
  X_x & \:=\; 
    \frac{\alpha}{v} \;.\,
    \left(\lambda\:+\:\frac{2}{3}\:\mu \right)
   \:.\:\tau
   \;-\; \lambda\:.\:\Theta
   \;-\; \frac{2}{3}\:.\:\mu\:.\:\Theta \; ,
  \nonumber  \\
  X_x & \:=\; 
    \left(\lambda\:+\:\frac{2}{3}\:\mu \right)
    \;.\,
    \left(\,\frac{\alpha\:\tau}{v}\:-\:\Theta\,\right)
  \;=\; Y_z \;=\; Z_x \; ,
  \;  \label{Eq_41} 
\end{align}
and, e.g. for $\Theta=0$, we find the pressure exerted by the body when it is heated at constant volume:
\vspace*{0mm}
\begin{align}
  X_x & \:=\; 
    \left(\lambda\:+\:\frac{2}{3}\:\mu \right)
    \;.\,
    \frac{\alpha\:\tau}{v}
  \; . \nonumber 
\end{align}

%
Furthermore, if the $X$-axis is assumed to be the direction of the tension, the following values are obtained for the elasticity tensor:  
\vspace*{-1mm} \begin{align}
 {\rm x}_x &\;=\;  a   \; ,
&{\rm y}_y &\;=\; -\:b \; ,
&{\rm z}_z &\;=\; -\:b \; ,
\nonumber \\
 {\rm x}_y &\:=\: 0 \; ,
&{\rm y}_z &\:=\: 0 \; ,
&{\rm z}_x &\:=\: 0 \; ,
 \; 
\nonumber \end{align} 
where $a$ denotes the dilation of the unit length and $2\:b$ the contraction of the unit cross-section.
From this follows: 
\vspace*{-2mm}
$$ \Theta \;=\; a \:-\; 2\:b $$ 
and from (\ref{Eq_40}) the values of the elastic forces: 
\vspace*{-1mm}
\begin{align}
 X_x &\;=\;
     \frac{\alpha}{v} \;.\,
     \left(\lambda\:+\:\frac{2}{3}\:\mu \right)
   \:.\:\tau
   \;-\; \left(\lambda\:+\:2\:\mu \right) \:.\: a
   \;+\; 2\:\lambda\:.\:b
 \;\; ,
\nonumber \\
 Y_y &\:=\: Z_z \;=\: 
     \frac{\alpha}{v} \;.\,
     \left(\lambda\:+\:\frac{2}{3}\:\mu \right)
   \:.\:\tau
   \;-\; \lambda \:.\: a
   \;+\; 2\:\left(\lambda\:+\:\mu\right)\:.\:b
 \;\; .
\nonumber
\end{align} 

If the surface conditions are taken into account, it follows that the elastic forces that do not act in the direction of the tension are $= 0$, and therefore: 
$$ \frac{\alpha}{v} \;.\,
     \left(\lambda\:+\:\frac{2}{3}\:\mu \right)
   \:.\:\tau
   \;-\; \lambda \:.\: a
   \;+\; 2\:\left(\lambda\:+\:\mu\right)\:.\:b
\; = \; 0 \; , $$
and hence :
\vspace*{-3mm}
\begin{align}
 2\;b &\;=\;
 \frac{\: 3 \: \lambda \:.\: v \:.\: a
   \; - \; 
  \left(\, 3 \: \lambda\:+\:2\:\mu \,\right)
   \:.\: \alpha \:.\: \tau \:}
 {3 \: \left(\,\lambda\:+\: \mu \,\right) \:.\: v}
 \;\; , \label{Eq_42}
\end{align} 
and consequently the elastic stress in the direction of the tension is: 
\vspace*{0mm}
\begin{align}
 X_x &\;=\; - \:
 \frac{\:\mu \:.\: 
   \left(\, 3 \: \lambda\:+\:2\:\mu \,\right)\:}
 {\left(\,\lambda\:+\: \mu \,\right)} \;\;
 \left(\, a \:-\: \frac{\alpha\:.\:\tau}{3\:\;v}\,\right)
 \;\; . \label{Eq_43}
\end{align} 
For $a=0$, for example, we obtain from this equation the stress that arises in the longitudinal direction of a rod when it is heated and its longitudinal expansion is prevented by external pressure, while the cross-section is allowed to expand. Then this pressure is: 
\vspace*{-1mm}
\begin{align}
 X_x &\;=\; 
 \frac{\:\mu \:.\: 
   \left(\, 3 \: \lambda\:+\:2\:\mu \,\right)\:}
 {3\;\left(\,\lambda\:+\: \mu \,\right)} \;\:.\:\;
 \frac{\alpha\:.\:\tau}{v}
 \;\; , \nonumber 
\end{align} 
and the corresponding dilatation of the cross-sectional unit according to (\ref{Eq_42}) is: 
\vspace*{-1mm}
\begin{align}
 -\:2\:b &\;=\; 
 \frac{\: 3 \: \lambda\:+\:2\:\mu \:}
 {3\;\left(\,\lambda\:+\: \mu \,\right)} \;\:.\:\;
 \frac{\alpha\:.\:\tau}{v}
 \;\; . \nonumber 
\end{align} 
From this, we arrive at the simple relationship: 
$$X_x \;=\; -\:2\:b \:.\:\mu \;\; .$$ 
It is also possible to introduce the conditions $b=0$ or $\Theta=0$.

If the body undergoes a change of state without heat being added or removed from the outside, its temperature will generally change. The condition applicable to this case is obtained by setting the entropy change, which is proportional to the heat supplied from outside, to $=0$. 
Therefore: 
\vspace*{-1mm}
\begin{equation}
  S {\color{red}\;-\; S_0}
  \;=\; \bigintssss d\,M \:.\:  
  (\, s {\color{red}\:-\: s_0} \,)
  \;=\; 0
  \; , \nonumber 
\end{equation}
or according to (\ref{Eq_38}): 
\vspace*{-1mm}
\begin{equation}
    \frac{k}{T} \:.\: \tau \:.\: M
  \;+\; 
  \alpha \:.\:
    \left( \lambda\:+\:\frac{2}{3}\;\mu \right)
  \:.\:  \bigintssss \, \Theta \:.\: \,dM
  \;=\; 0
  \; . \nonumber 
\end{equation}
If $\Theta$ is equivalent for all points on the body, the equation simplifies to: 
\vspace*{0mm}
\begin{equation}
    \frac{k}{T} \:.\: \tau 
  \;+\; 
  \alpha \:.\:
    \left( \lambda\:+\:\frac{2}{3}\;\mu \right)
  \:.\:  \Theta 
  \;=\; 0
  \; . \label{Eq_44}
\end{equation}
If, for example, we consider the case of cubic compression, assuming that no heat is released to the outside, and look for the temperature increase caused by the compression, equations (\ref{Eq_41}) and (\ref{Eq_44}) apply.

From them, if the specific heat $c$ at constant pressure is also introduced by means of (\ref{Eq_39}) for simplification, the temperature increase is: 
\vspace*{-3mm}
\begin{align}
 \tau &\;=\; \frac{\alpha\:.\:T}{c} \:\:.\:\: X_x
 \;\; , \nonumber 
\end{align} 
and furthermore, the compression of the volume unit: 
\vspace*{0mm}
\begin{align}
 \Theta & \;=\; -\:\frac{k}{c} \:\:.\:\:
  \frac{3}{\: 3 \: \lambda\:+\:2\:\mu \:} 
  \:\:.\:\: X_x
 \;\; . \nonumber 
\end{align} 

The laws that apply to tension elasticity arise analogously if no heat is supplied from outside. 
In this case, equations (\ref{Eq_42}), (\ref{Eq_43}) and (\ref{Eq_44}) apply, from which the temperature change is obtained using equation (\ref{Eq_39}): 
\vspace*{-3mm}
\begin{align}
 \tau &\;=\; \frac{\alpha\:.\:T}{3\:\:c} \:\:.\:\: X_x
 \;\; , \nonumber 
\end{align} 
where $X_x$ is negative in the case of tension.

Furthermore, the length dilation is: 
\vspace*{0mm}
\begin{align}
 a & \;=\; 
 -\:\frac{1}{3} \:
  \left(\,
  \frac{1}{\mu} 
  \;+\;
  \frac{k}{c} \;\:
  \frac{1}{\: 3 \: \lambda\:+\:2\:\mu \:} 
  \,\right)
  \: X_x
 \;\; , \nonumber 
\end{align} 
and the transverse contraction is: 
\vspace*{0mm}
\begin{align}
 2\;b & \;=\; 
 -\:\frac{1}{3} \:
  \left(\,
  \frac{1}{\mu} 
  \;-\;
  \frac{k}{c} \;\:
  \frac{2}{\: 3 \: \lambda\:+\:2\:\mu \:} 
  \,\right)
  \: X_x
 \;\; , \nonumber 
\end{align} 
Subtracting the last two equations gives the volume dilatation: 
\vspace*{0mm}
\begin{align}
 \Theta & \;=\; a \:-\: 2 \; b \;=\; 
 -\:\frac{k}{c} \:\:.\:\:
  \frac{X_x}{\: 3 \: \lambda\:+\:2\:\mu \:} 
 \;\; , \nonumber 
\end{align} 
and dividing them gives the ratio of the transverse contraction to the length dilatation: 
\vspace*{0mm}
\begin{align}
 \frac{2\;b}{a} & \;=\; 
  \frac{(\,3\:\lambda\:+\:2\:\mu\,)\;c\;-\;2\;\mu\;\,k}
       {(\,3\:\lambda\:+\:2\:\mu\,)\;c\;+\;\mu\;\,k} 
 \;\; . \nonumber 
\end{align} 

The greater the difference between the two specific 
heats {\it(namely $c$ and $k$)},
the more significantly the quantities calculated here deviate from those which express the conditions of tension elasticity at constant temperature, 
which are obtained directly by substituting $\tau=0$ in equations (\ref{Eq_42}) and (\ref{Eq_43}).

\subsubsection{\underline{Pure Liquid drop} (p.27-28)}
\vspace*{-1mm}

  For pure liquid drops, 
the expressions of the elastic forces are simplified in that in equilibrium states they always act only perpendicularly on a surface element. 
We therefore have in general: 
$$ X_y \;=\; 0 \;=\; Y_z \;=\; Z_x \; , $$ 
and consequently according to (\ref{Eq_32}): 
$$ \mu \;=\; 0 \; . $$
If moreover we again assume an initial pressure of $P=0$, then according to (\ref{Eq_36}): 
\vspace*{0mm}
\begin{align}
  X_x & \:=\; 
  \lambda \:.
   \left(\:
   \frac{\alpha}{v}\:\:\tau
   \;-\; 
   \Theta
   \:\right)
  \;=\; Y_y \;=\; Z_z
  \;\; . \label{Eq_45} 
\end{align}

The value of $X_x$ must then be the same at every point of the body according to the equilibrium conditions inside (\ref{Eq_1}), and consequently the volume dilation $\Theta$ is also equivalent for all points.

Then the expression of the entropy is simplified according to (\ref{Eq_38})
to:$\,$\footnote{\color{red}$\:${\it As explained in the footnote~\ref{footnote_s0_bis} about the equation~(\ref{Eq_38}), I have written in red the alternative formulation obtained with $\tau/T=(T'-T)/T=\Delta T/T=\Delta \ln(T)$ and similarly with $\Theta=(v'-v)/v=\Delta v/v=\Delta \ln(v)$.
I have also added, as explained in the footnote~\ref{footnote_S0},
the (global) reference value {\color{red}$S_0 = s_0 \:M$}.
The formulation (\ref{Eq_46}) of Planck for {\color{red}$\Delta S = S - S_0$} then looks like the modern formulation for the entropy, with {\color{red}$M\:k$} corresponding to the ``specific heat at constant volume'' $C_v=M\:c_v$, and with the second bracketed term {\color{red}$[\:M\:\alpha\:\lambda\:]$} driving the impact of a change of volume (P. Marquet)}.}  
\vspace*{0mm}
\begin{equation}
  S {\color{red}\;-\; S_0} 
  \;\approx\; 
  \frac{k}{T} \; \tau \; M
  \;+\; 
  \alpha \; \lambda \; \Theta \; M
  {\color{red}\;\;\;\;\;
  \mbox{or alternatively:}\;\; \Delta S \;\approx\; 
   [\:M\:k\:] \; \Delta\ln(T)
  \;+\; 
  [\:M\:\alpha\:\lambda\:] \;\: \Delta\ln(v)
  }
  \; , \label{Eq_46}
\end{equation}
and the difference of the specific heats (\ref{Eq_39}) 
becomes:$\,$\footnote{\color{red}$\:${\it As explained in the footnote~\ref{footnote_c_k_T}, (\ref{Eq_47}) corresponds to a specific heat given by a linear (affine) function of the absolute temperature: $c \:=\: k \:+\: [\:{\alpha^2 \; \lambda\:}/\,{v}\:]\:T$ (P. Marquet)}.}
\begin{equation}
  c \:-\: k 
  \;=\; 
  \frac{\alpha^2 \; \lambda}{v}
  \;\; T
  \; . \label{Eq_47}
\end{equation}
The physical meaning of $\lambda$ results from (\ref{Eq_45}) and if one sets $\tau=0$: 
\vspace*{0mm}
\begin{align}
  X_x & \:=\; 
   -\:\lambda \:.\: \Theta
  {\color{red}\;\;\;=\; Y_y \;=\; Z_z}
  \;\; , \nonumber 
\end{align}
and $\lambda$ thus represents the reciprocal value of the compression coefficient at constant temperature.
Furthermore, for $\Theta=0$, the pressure exerted by the liquid when it is heated at constant volume results from (\ref{Eq_45}): 
\vspace*{-3mm}
\begin{align}
  X_x & \:=\; 
  \frac{\alpha \; \lambda}{v} \:\:.\:\: \tau
  {\color{red}\;\;\;=\; Y_y \;=\; Z_z}
  \;\; . \nonumber 
\end{align}

If no heat is released to the outside during compression, the entropy change {\color{red}$\Delta S\:$}$=0$, i.e. according to (\ref{Eq_46}): 
$$ 
  \frac{k\;\tau}{T} 
  \;+\; 
  \alpha \; \lambda \; \Theta
  \;=\; 0 
  {\color{red}\;\;\;\;\;
  \mbox{or alternatively:}\;\; 
   k \; \Delta\ln(T)
  \;+\; 
  [\:\alpha\:\lambda\:] \;\: \Delta\ln(v)
  \;=\; 0 
  }
  \;\;.
$$
From this, and from (\ref{Eq_45}), the temperature change caused by the compression can be found by introducing $c$ from (\ref{Eq_47}): 
\vspace*{-2mm}
$$ 
  \tau \;=\; \frac{\alpha \; T}{c} \;.\; X_x \;\;,
$$ 
and the compression of the volume unit is: 
\vspace*{-2mm}
$$ 
  \Theta \;=\; -\:\frac{k}{c} \;.\; \frac{X_x}{\lambda} \;\;.
$$ 
\vspace*{-2mm}

\subsubsection{\underline{Vapours and gases} (p.28-32)}
\vspace*{-1mm}

For these cases, the formulae change insofar as the pressure $P$ corresponding to the initial state cannot be assumed to be $=0$, as is the case with liquids and solids.  
Otherwise, the same formulae apply as for pure liquids, in that $\mu=0$ again. 
We therefore have, according to (\ref{Eq_36}): 
\vspace*{0mm}
\begin{align}
  X_x \;=\; Y_y \;=\; Z_z
 & \:=\; 
  P \;+\;
  \lambda \:.
   \left(\:
   \frac{\alpha}{v}\:\:\tau
   \;-\; 
   \Theta
   \:\right)
  \;\; . \nonumber 
\end{align}
Furthermore, the specific entropy according to (\ref{Eq_38}):
\vspace*{-1mm}
\begin{equation}
  s {\color{red}\;-\; s_0} 
  \;\approx\; 
  \frac{k}{T} \:.\: \tau 
  \;+\;
  \alpha \:.\: \lambda \:.\: \Theta 
  \; ,
  {\color{red}\;\;\;\;\;\;\;
  \mbox{or equivalently:}\;\; \Delta s \;\approx\; 
   k\;\: \Delta\ln(T)
  \;+\; 
  \left[\:
  \alpha \:.\: \lambda 
  \:\right]
  \: \Delta\ln(v) 
  }
  \; , \nonumber  
\end{equation}
and the specific energy according to (\ref{Eq_37}) is: 
\vspace*{-2mm}
\begin{align}
  u {\color{red}\;-\; u_0} 
  &\;\approx\; k \:.\: \tau 
  \;+\; 
  \left\{\:
  \alpha \:.\: \lambda \:.\: T
    \;-\; P \:.\: v
  \:\right\}
  \:.\: \Theta 
  \; , \nonumber 
  {\color{red}
  \;\;\mbox{or equivalently:}\;\;
  \Delta u \;
  \;\approx\; k \; \Delta T 
  \;-\; 
  \left\{\:
     P 
    \;-\; 
    \alpha \:.\: \lambda \:.\: \frac{T}{v}
  \:\right\}
  \; \Delta v
  }
  \; . \nonumber
\end{align}
Finally, according to (\ref{Eq_39}), the difference between the two specific heats
is:$\,$\footnote{\color{red}\label{footnote_c_k_Tbis}$\:${\it Namely, as explained in the footnote~\ref{footnote_c_k_T}, a linear (affine) function of the absolute temperature $T$ around $c_v(T_0)= k + T_0\: \alpha^2\: \lambda \,/\,v$ (P. Marquet)}.}
\begin{equation}
  c \:-\: k 
  \; \approx \; 
  \frac{\alpha^2\:\lambda\:T}{v}
  \; ,
  {\color{red}
  \;\;\;\;\;\;\mbox{or equivalently:}\;\;\;
  c_v\,(T) \;
  \;\approx\; 
  c_v\,(T_0)
  \;+\; 
  \left[\:
     \frac{\alpha^2\:\lambda}{v}
  \:\right]
  \; (\:T-T_0\:)
  }
  \; . \nonumber 
\end{equation}
Of course, these equations apply regardless of how a vapour or gas behaves with respect to Mariotte's and Gay Lussac's law. Therefore, their range of validity only extends to very small changes of 
state.$\,$\footnote{\color{red}$\:${\it This corresponds to the alternative formulations for $\Delta u$ and $\Delta s$ derived with small values of $\tau=T'-T=\Delta T$ and $\Theta = (v'-v)/v=\Delta v/v$. This is also in agreement with the linear (affine) formulation for the specific heat $c_v(T)$, considered as first-order variations around a reference value $c_v(T_0)\:=\:k + T_0\: \alpha^2\: \lambda \,/\,v$ and with ambient values of $T$ remaining close to $T_0$ (P. Marquet)}.}

The applications of these formulae to the behaviour of a gaseous body at constant volume, constant pressure, constant temperature, constant entropy or constant energy (expansion without overcoming external pressure) are immediate.

\vspace*{2mm}
\begin{center}
--------------------------------------------------- 
\end{center}
\vspace*{-2mm}

Up to now we have only considered very small changes of state.
However, the investigations we have made can also be extended in certain respects to changes of state of any size, if we consider the initial state of the body under consideration to be variable.

According to the assumptions made at the beginning of this section, this initial state is 
subject to no other condition than: that the body 
is completely homogeneous in it; 
and that a normal pressure that is the same everywhere and of any size acts on its surface.
From this it follows, as we have seen from equation (\ref{Eq_29}), that in the initial state the internal forces of the body are also reduced to a pressure which is the same everywhere, acting perpendicularly on each surface element, and which is equal and opposite to the external pressure.

For a given body, we therefore have the choice between an infinite number of states, each of which we can use as the initial state for the derived equations. 
Let us now try to determine by which quantities such a state, which can serve as the initial state, is completely determined.

In the introduction we stated the premise that the state of an isotropic body of a certain chemical composition and a certain mass is completely determined if we know: 1) its temperature; 2) the position of all its smallest particles. 
From this it can be concluded that such a state as we have to consider here, in which the body is completely homogeneous and an elastic force acting in all directions, is completely determined by the value of: 1) the temperature; 2) the volume occupied by the unit of mass (because the position of all the mass particles is given by the latter, and  because in this case they fill the body in all directions in a completely uniform manner).
\vspace*{2mm}

We can therefore assume that the initial state for which the derived equations apply is determined if the temperature $T$ and the specific volume $v$ of an isotropic body, with a certain chemical composition and a certain mass, are given.

Therefore, for a given body all the constants occurring in the above equations (as the initial pressure $P$, the coefficients of elasticity $\lambda$ and $\mu$, the coefficient of thermal expansion $\alpha$, the specific heats $c$ and $k$), since they depend only on the chosen initial state, are functions of $T$ and $v$, and we will now draw the conclusions that result from those equations using this fact.

If we change the initial state infinitely little by allowing $T$ and $v$ to become $T+dT$ and $v+dv$, we can apply all the equations derived above to this change of state, because it is infinitely small, and thus obtain the corresponding change in pressure, energy and entropy. 
The following must be set: 
\vspace*{-1mm}
\begin{align}
   \tau & \:=\; d\,T \; ,
&  \Theta & \:=\; \frac{d\,v}{v}  \; ,
\nonumber \\
\mbox{and finally}\;\;\;\;
   X_x   \;=\; Y_y \;=\; Z_z & \:=\; P \:+\: d\, P  \; ,
&  X_y & \:=\; Y_z \;=\; Z_x \;=\; 0  \; .
\nonumber 
\end{align}

Since the body should have the properties of an initial state again after the change of state, the pressure inside it must again be the same everywhere and directed vertically onto each surface element.

If we first insert these values into equations (\ref{Eq_36}) and (\ref{Eq_32}), we obtain: 
\begin{align}
  X_x \;=\; Y_y \;=\; Z_z 
  & 
  \;\;\;\mbox{i.e.} \;\;
  \:=\; \frac{\Theta}{3} 
  \;=\; \frac{dv}{3\;v} \; ,
\nonumber \\
  X_y \;=\; Y_z \;=\; Z_x    & \:=\; 0  \; ,
\nonumber 
\end{align}
and then :
\vspace*{-3mm}
\begin{align}
  dP & \:=\; 
  \frac{\alpha}{v} \:.
  \left(\,
  \lambda \;+\; \frac{2}{3}\:\mu
  \,\right)
  dT
  \;-\; 
  \frac{1}{v}  \:.
  \left(\,
  \lambda \;+\; \frac{2}{3}\:\mu
  \,\right)
  dv
\; . \nonumber 
\end{align}
If we denote the partial derivatives 
of $P$ with respect to $T$ and $v$ with $\partial P / \partial T$ and $\partial P / \partial v$, this results in:
\vspace*{-3mm}
\begin{align}
 \frac{\partial P}{\partial\,T} 
   & \:=\; 
  \frac{\alpha}{v} \:.
  \left(\,
  \lambda \;+\; \frac{2}{3}\:\mu
  \,\right)
\; , \label{Eq_48} \\  
 \frac{\partial P}{\partial v} 
   & \:=\; -\; 
  \frac{1}{v}  \:.
  \left(\,
  \lambda \;+\; \frac{2}{3}\:\mu
  \,\right)
\; , \nonumber 
\end{align}
and consequently: 
\vspace*{-3mm}
\begin{align}
 \frac{\partial P}{\partial\,T} 
   & \:=\; 
 -\:\alpha \:.\:
 \frac{\partial P}{\partial v} 
\; . \label{Eq_49} 
\end{align}
Furthermore, we obtain from (\ref{Eq_37}) the corresponding change of the specific energy $u$: 
\vspace*{0mm}
\begin{align}
 du & \:=\;
  k \:\: dT
  \;+\;
  \left[\:
  \alpha \;
  \left(\,
  \lambda \;+\; \frac{2}{3}\:\mu
  \,\right)
  \: T 
  \;-\; P\:v
  \:\right]
  .\:
  \frac{dv}{v}
\; ,  \nonumber 
\end{align}
or according to (\ref{Eq_48}): 
\vspace*{-3mm}
\begin{align}
 du & \:=\;
  k \:\: dT
  \;+\;
  \left(\:
    T \:.\: \frac{\partial P}{\partial\,T} \;-\; P
  \:\right)
  \: dv
\; ,  \nonumber 
\end{align}
and hence: 
\vspace*{-3mm}
\begin{align}
 \frac{\partial u}{\partial T} 
 & \:=\; k
\; , \label{Eq_50} 
\\
 \frac{\partial u}{\partial v} 
 & \:=\; 
    T \:.\: \frac{\partial P}{\partial\,T} \;-\; P
\; . \label{Eq_51} 
\end{align}

If we differentiate these two equations 
according to $V$ and $T$, and equating the two expressions
${\partial^2 u}/({\partial T}\:{\partial v})$
and  
${\partial^2 u}/({\partial v}\:{\partial T})$
obtained in this way equal to each other, 
we obtain: 
\vspace*{0mm}
\begin{align}
 \frac{\partial k}{\partial v} 
 & \:=\; T \:.\:\, \frac{\partial^2 P}{\partial\,T^2} 
\; . \label{Eq_52} 
\end{align}
Finally, for the specific entropy one has by (\ref{Eq_38}) and using (\ref{Eq_48}): 
\vspace*{0mm}
\begin{align}
 ds & \:=\;
  \frac{k}{T} \:\: dT
  \;+\;
  \frac{\partial P}{\partial\,T} 
  \:\: dv
\; ,  \label{Eq_53} 
\end{align}
so that: 
\vspace*{-3mm}
\begin{align}
 \frac{\partial s}{\partial T} 
 & \:=\; \frac{k}{T}
\; , \nonumber 
\\
 \frac{\partial s}{\partial v} 
 & \:=\; \frac{\partial P}{\partial\,T} 
\; . \nonumber 
\end{align}
By comparing the expressions for $du$ and $ds$, one immediately obtains: 
\vspace*{0mm}
\begin{align}
 ds
 & \:=\; 
 \frac{du \;+\; P\:.\;dv}{T}
\; . \label{Eq_54} 
\end{align}
If we add the equation (\ref{Eq_39}) that applies to the two specific heats, as it appears using equation (\ref{Eq_48}): 
\vspace*{0mm}
\begin{align}
 c \;-\; k 
 & \:=\; 
 \alpha \;\: T \;.\;\: \frac{\partial P}{\partial\,T} 
\; , \label{Eq_55} 
\end{align}
then we have exactly the equations that apply in mechanical heat theory for a homogeneous body determined by temperature and volume in its state, which do not give rise to any further conclusions here, and will only be applied several times in the next section.

\vspace*{2mm}
\begin{center}
--------------------------------------------------- 
\end{center}
\vspace*{-2mm}

\section{\underline{Sufficient equilibrium conditions} (p.33-63)} 
\vspace*{-2mm}

We have previously developed the conditions that must necessarily apply to an isotropic body when it is in equilibrium. These conditions were derived from the principle that in any state of equilibrium of any body:
\vspace*{-2mm}
\begin{enumerate}
\vspace*{-1mm}
\item \!\!\!) 
the temperature throughout the body is the same everywhere; and
\item \!\!\!) 
each element of the body, if considered rigid, is kept in equilibrium by the forces acting on its lateral surfaces. 
\end{enumerate}
\vspace*{-2mm}
The two main theorems of mechanical heat theory were applied to these conditions.

If we confine ourselves, as we shall always do in the following, to the consideration of such states of an isotropic body (in which the forces acting from outside on the surface are everywhere equal and directed normally towards it), then the second of the two necessary equilibrium conditions stated is reduced to the fact that the forces acting inside the body consist in a pressure which is everywhere equal and opposite to the external pressure, and which acts vertically on each surface element.

But if we now examine whether these conditions are also sufficient for equilibrium, i.e. if, conversely, we ask the question: if an isotropic body is in a state in which its temperature is the same everywhere and an equal pressure perpendicular to each surface element acts everywhere in its interior (which is equal and opposite to the external pressure), is this state then always a state of equilibrium? --\,experience teaches us that this question must be answered in the negative.

Because, for example, it is well known that there is an infinite number of vapour states 
which cannot exist in a homogeneous form, 
whereas it must be assumed that, in every case where a vapour is completely homogeneous and its temperature is the same everywhere, the pressure inside it also acts completely uniformly in all directions.
However, if one attempts to bring a vapour into such a (supersaturated) state, i.e. if one assigns it a corresponding temperature and a corresponding volume, it does not take on a homogeneous form, but rather partly precipitates in the form of another aggregate state.

The same applies to liquids and solids: there are also states in which they are completely homogeneous and of uniform temperature, but are nevertheless not in equilibrium, but partly vaporise or melt.

Another remarkable fact here is that cases occur where a body in homogeneous form is in equilibrium, but where this equilibrium is only an unstable one and is disturbed by the slightest external cause, which phenomenon is observed, for example, in water when it is allowed to cool below 0°C at normal pressure.

We thus see that the uniformity of temperature and pressure inside a body is by no means sufficient for stable equilibrium, but that this equilibrium is sometimes only an unstable one, and sometimes does not exist at all. 
Therefore, in the following, we will look at the conditions that are completely sufficient for equilibrium, and also establish the distinguishing characteristics of stable and unstable equilibrium.

What we presuppose is only this: that every state of a certain isotropic body, in which it is completely homogeneous and has a uniform pressure and temperature distribution in all directions, whether it is a state of equilibrium or not, is completely determined by the mass, the temperature and the specific volume, and that the other determinants of the state, such as pressure, energy, entropy, are certain unambiguous functions of these quantities.

The present problem can be most conveniently formulated as follows: we imagine a body of a certain mass enclosed in a certain volume, and a certain energy imparted to it (perhaps by the imparting or withdrawing of heat).
We then seek to assume the body (or if there are several bodies), that can be subject to the given circumstances, and at the same time to indicate the conditions under which the equilibrium is stable or unstable.

The realisation of this investigation is made possible by the application of the principle which forms the most general version of the second law of the mechanical theory of heat.
According to the formulation and detailed justification I have attempted, it reads as follows: \\
If one imagines a group of bodies in any two states, and designates the sum of the entropies of the bodies in the first state as $S_1$ and in the second state as $S_2$, then, if $S_2>S_1$, a process is always possible in nature which transfers all the bodies from the first state exactly into the second, while no process is possible in the opposite direction. 
It is assumed that the two sums of entropies extend to all bodies that are in both states under somehow different circumstances.

According to Clausius, this sentence can also be expressed briefly as follows: Nature strives 
to increase the sum of the entropies of all bodies.

In order to be able to apply this principle to the present case, we must examine and compare the different values which the entropy of the body under consideration assumes, when it is in all conceivable states with the given total volume and with the given total energy. 
The state to which the greatest value of entropy corresponds will represent a state of equilibrium (namely a stable one), because once this state has been achieved, as long as the external conditions are maintained (i.e. the volume and the energy are the same), no change in nature can occur because, according to the principle stated above, such a change would necessarily bring with it an increase in entropy, but this is not possible according to the assumptions made.

\begin{figure}[hbt]
\centering
\vspace*{2mm} \hrule  \vspace*{2mm} 
\includegraphics[width=0.8\linewidth]{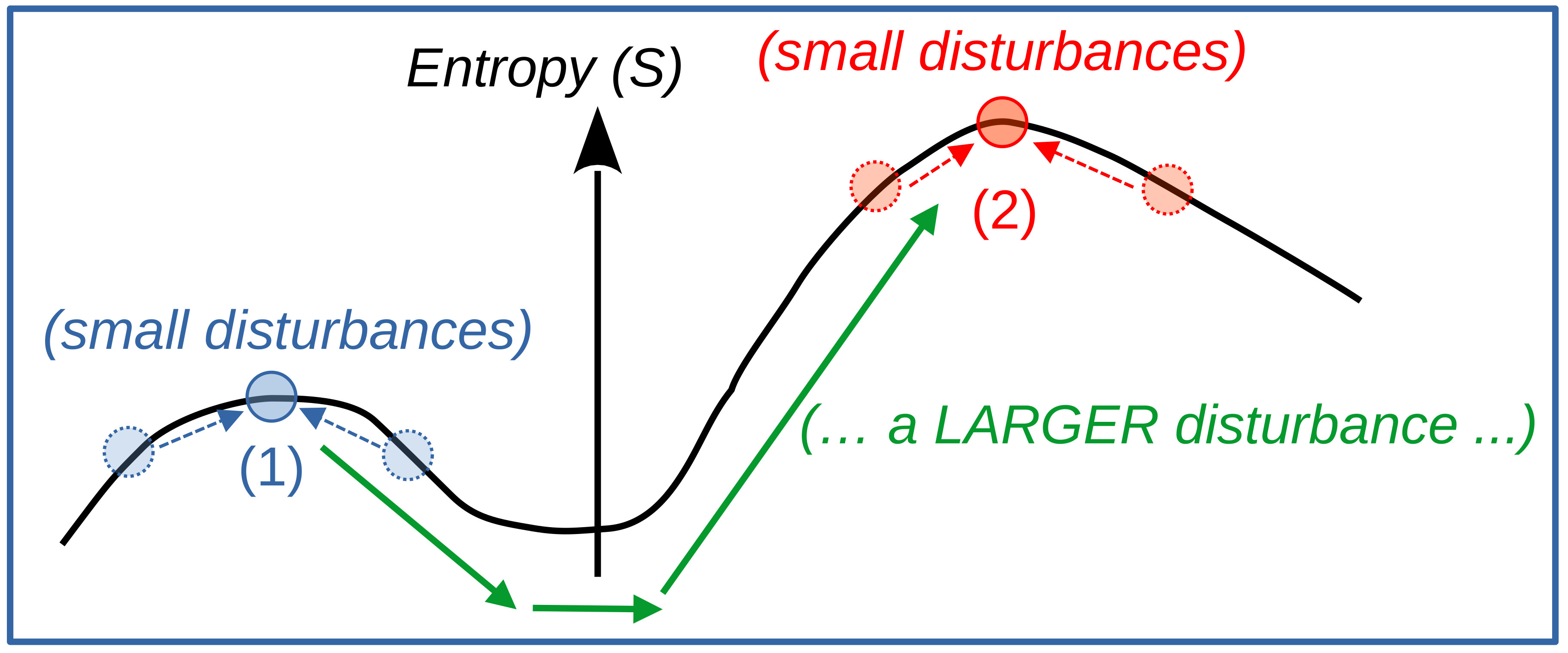} 
\vspace{-3mm}
\caption{\color{red}\it An illustration of the multiple maxima of the entropy described by \citet[][]{Planck_Habilitation_thesis_1880}: (1) a relative maximum state (in blue) stable for small disturbances; and (2) the absolute maximum state (in red) reachable from (1) via larger (green) disturbances (figure plotted by P. Marquet).
\label{fig_Planck_1880_maximas}}
\vspace*{2mm} \hrule  \vspace*{2mm}
\end{figure}

But it may also be the case that the entropy of the body under the given external conditions may assume several relative maxima.
Then each relative maximum, which is not the absolute one, corresponds to an unstable state of equilibrium, because if the body is in such a state, it will, if only very small changes of state are taken into consideration, have to remain in this state for the reason given above, but if a sufficiently large disturbance occurs from outside, the body may under certain circumstances be permanently removed from the state and pass into another state of equilibrium, 
which corresponds to a greater value of entropy than the previous one
{\it\color{red}(see the Fig.\ref{fig_Planck_1880_maximas})}.

In any case there must be an absolute maximum of entropy, because if entropy could grow to infinity, a stable state of equilibrium would never occur, which contradicts experience.

We must now first seek out those states in which the entropy of the body reaches a maximum.
This is done by establishing the necessary condition which such a state must fulfil. 
If $S$ is the entropy of the body in such a state, which we can briefly call the maximum state, then for every infinitesimal change of state compatible with the given external conditions, the variation of entropy must be: $\delta S = 0$. 
Among the states that fulfil the latter condition are in any case  the desired maximum states, and thus also the state of the absolute maximum, i.e. the stable equilibrium
{\it\color{red}(namely the state (2) in the Fig.\ref{fig_Planck_1880_maximas})}.

The most general form under which such a maximum state, which, as we have seen, is always also a state of equilibrium, 
 can appear is that different parts of the body are in different states of aggregation.
If we therefore denote the quantities of the body in the $3$ different states of aggregation by $M_1$, $M_2$ and $M_3$ 
(where the specific meaning of the individual indices is left open),
we have, if $M$ is the given mass of the whole 
body:$\,$\footnote{\color{red}\label{footnote_sum_k}$\:${\it Here and in the following I have replaced the old-fashioned notations of Planck like ``\:$\sum\,M_1$'' by the Einstein summation over the $k$ indices, like: ``\:$\sum_k\,M_k$'' (P. Marquet)}.}
\vspace*{-3mm}
\begin{align}
 M & \:=\; M_1 \;+\;  M_2 \;+\;  M_3
  \;=\; \sum_k \: M_k
\; . \label{Eq_56} 
\end{align}
\vspace*{-5mm}

\noindent Since the state considered here is a state of equilibrium, each of these $3$ parts of the body must also be in equilibrium, i.e. of uniform temperature and homogeneous, because it must be regarded as a conclusion from experience that equilibrium 
is not possible in any other way between parts of the same aggregate state.

If $V$ denotes the given volume of the body, then we have: 
\vspace*{0mm}
\begin{align}
 V & \:=\; M_1\:\:v_1 \;+\;  M_2\:\:v_2 \;+\;  M_3\:\:v_3
  \;=\; \sum_k \: M_k \:\: v_k
\; , \label{Eq_57} 
\end{align}
\vspace*{-5mm}

\noindent where $v_1$, $v_2$, $v_3$ denote the specific volumes of the $3$ parts of the body.

Similarly, one obtains for $U$, the given energy of the body: 
\vspace*{0mm}
\begin{align}
 U & \:=\; M_1\:\:u_1 \;+\;  M_2\:\:u_2 \;+\;  M_3\:\:u_3
  \;=\; \sum_k \: M_k \:\: u_k
\; , \label{Eq_58} 
\end{align}
\vspace*{-5mm}

\noindent where $u$ again denotes the energy of the unit of mass (specific energy).

The three equations (\ref{Eq_56}), (\ref{Eq_57}) and (\ref{Eq_58}) correspond to the given external conditions.

For the entropy we now obtain: 
\vspace*{0mm}
\begin{align}
 S & \:=\; M_1\:\:s_1 \;+\;  M_2\:\:s_2 \;+\;  M_3\:\:s_3
  \;=\; \sum_k \: M_k \:\: s_k
\; , \nonumber 
\end{align}
\vspace*{-5mm}

\noindent where $s$ is the specific entropy.

Therefore the following equation results for an infinitely small change of state:
\vspace*{0mm}
\begin{align}
 \delta\,S &
  \;=\; \sum_k \: M_k \:.\: \delta\,s_k
  \;+\; \sum_k \: \delta\,s_k \:.\: M_k
\; , \nonumber 
\end{align}
\vspace*{-5mm}

\noindent 
and taking into account that according to (\ref{Eq_54}) in general: 
\vspace*{0mm}
\begin{align}
 \delta\,s
 & \:=\; 
 \frac{\delta\,u \;+\; P\:.\;\delta\,v}{T}
\; , \label{Eq_59} 
\end{align}
one obtains by substitution of this value:
\vspace*{0mm}
\begin{align}
 \delta\,S &  
  \;=\; \sum_k \: \frac{M_k \:.\: \delta\,u_k}{T_k}
  \;+\; \sum_k \: \frac{M_k \:.\: P_k\:.\: \delta\,v_k}{T_k}
  \;+\; \sum_k \: s_k \:.\: \delta\,M_k
\; . \label{Eq_60} 
\end{align}
\vspace*{-5mm}

\noindent 
However, the variations of the individual determinants of the change of state are not completely independent of each other, 
rather, by variation it follows from equations (\ref{Eq_56}), (\ref{Eq_57}) and (\ref{Eq_58}):
\vspace*{0mm}
\begin{align} 
  \sum_k \: \delta\,M_k \;=\; 0
\; , \hspace*{20mm} \label{Eq_61}  \\
  \sum_k \: M_k \:.\: \delta\,v_k
  \;+\;
  \sum_k \: v_k \:.\: \delta\,M_k \;=\; 0
\; , \label{Eq_62}  \\
  \sum_k \: M_k \:.\: \delta\,u_k
  \;+\;
  \sum_k \: u_k \:.\: \delta\,M_k \;=\; 0
\; . \label{Eq_63} 
\end{align}
\vspace*{-5mm}

With the help of these $3$ equations, we must therefore eliminate any $3$ variations from the expression of $\delta\,S$ in order to obtain only variations that are completely independent of each other. 
If, for example, we calculate the values of $\delta\,M_2$, $\delta\,v_2$ and $\delta\,u_2$ from these last equations and insert them into (\ref{Eq_60}), we obtain: 
\vspace*{-3mm}
\begin{align}
 \delta\,S & \:=\; 
  \left(\, 
    \frac{1}{T_1}\:-\:\frac{1}{T_2}
  \,\right)
  .\: M_1 \:.\: \delta\,u_1
  \;-\;
  \left(\, 
    \frac{1}{T_2}\:-\:\frac{1}{T_3}
  \,\right)
  .\: M_3 \:.\: \delta\,u_3
\nonumber \\
  & \:+\; 
  \left(\, 
    \frac{P_1}{T_1}\:-\:\frac{P_2}{T_2}
  \,\right)
  .\: M_1 \:.\: \delta\,v_1
  \;-\;
  \left(\, 
    \frac{P_2}{T_2}\:-\:\frac{P_3}{T_3}
  \,\right)
  .\: M_3 \:.\: \delta\,v_3
\nonumber \\
  & \:+\; 
  \left(\, 
   s_1\:-\:s_2
   \;-\;
   \frac{u_1\:-\:u_2}{T_2}
   \;-\;
   \frac{P_2\:(\,v_1\:-\:v_2\,)}{T_2}
  \,\right)
  .\: \delta M_1
\nonumber \\
  & \:-\; 
  \left(\, 
   s_2\:-\:s_3
   \;-\;
   \frac{u_2\:-\:u_3}{T_2}
   \;-\;
   \frac{P_2\:(\,v_2\:-\:v_3\,)}{T_2}
  \,\right)
  .\: \delta M_2
\;\: . \label{Eq_64} 
\end{align}

Since the $6$ variations contained in this expression are completely independent of each other, all six coefficients of this variation must disappear so that $\delta\,S$ is $=0$ for all possible changes of state. Thus we have: 
\vspace*{-3mm}
\begin{align}
  & \hspace*{5mm} T_1 \;=\; T_2 \;=\; T_3
\;\: , \label{Eq_65} 
 \\
  & \hspace*{5mm} P_1 \;=\; P_2 \;=\; P_3
\;\: , \label{Eq_66} 
 \\
  s_1\:-\:s_2 & \:=\;
  \frac{(\,u_1\:-\:u_2\,) \:+\: P_1\:(\,v_1\:-\:v_2\,)}{T_1}
\;\: , \label{Eq_67} 
 \\
  s_2\:-\:s_3 & \:=\;
  \frac{(\,u_2\:-\:u_3\,) \:+\: P_1\:(\,v_2\:-\:v_3\,)}{T_1}
\;\: . \label{Eq_68} 
\end{align}

These $6$ equations represent the necessary properties of a state to which a maximum of entropy corresponds, i.e. a state of equilibrium. The first $4$ of them 
{\color{red}{\it--\,namely $2$ in (\ref{Eq_65}) and $2$ in (\ref{Eq_66})\,--}}
express the equality of temperature and pressure, which was already established in the previous section as a necessary condition of equilibrium. The main interest is therefore concentrated on the last two equations 
{\color{red}{\it--\,namely (\ref{Eq_67}) and (\ref{Eq_68}) for the difference in entropies\,--}}, each of which contains the entire existing theory of saturated vapours and the melting and evaporation process.

Let us first bring these {\color{red}{\it(last)}} two equations to a somewhat simpler form by substituting their value for the specific entropy $s$, which can be regarded as a function of $T$ and $v$.

Since in general according to (\ref{Eq_54}): 
\vspace*{-3mm}
\begin{align}
 ds
 & \:=\; 
 \frac{du \;+\; P\:.\;dv}{T}
\; , \nonumber 
\end{align}
we have, if we integrate this equation and apply it to our case: 
\vspace*{0mm}
\begin{align}
 s_1\:-\:s_2 
 & \:=\; 
 \bigintsss_{\,2}^1 
 \frac{du \;+\; P\:.\;dv}{T}
\;\: . \nonumber 
\end{align}
Here, the upper limit of the integral is completely determined by the pair of values $T=T_1$ and $v=v_1$, and the lower limit by the pair of values $T=T_2$ and $v=v_2$.

However, since according to (\ref{Eq_65}) $T_1=T_2$, we can leave the temperature constant {\color{red}{\it(equal to $T_1$)}}  when carrying out the integration, and thus obtain: 
\vspace*{-1mm}
\begin{align}
 s_1\:-\:s_2 
 & \:=\; 
 \frac{1}{T_1} \:.\: (\,u_1\:-\:u_2\,) 
   \;+\;
  \frac{1}{T_1} \:.\! \bigintsss_{\,v_2}^{v_1} 
  \!\! P \; \partial\,v
\;\: . \nonumber 
\end{align}
Here, the sign $\partial\,v$ (with round $\partial$) means that the integration is to be carried out partially according to $v$, whereas $T$, the other variable on which $P$ still depends, remains constant.

If we now substitute this value of $(s_1-s_2)$ into the equation (\ref{Eq_67}), we obtain the relation: 
\vspace*{0mm} 
\begin{equation}
\left.
\begin{aligned}
  &
 \hspace*{5mm} 
  \bigintsss_{\,v_2}^{v_1}\!\! P\;\partial\,v
  \; = \; 
  P_1 \:.\: (\,v_1\:-\:v_2\,)
  \;\: ,  \\
 \mbox{with also from (\ref{Eq_68}):} &
 \hspace*{5mm} 
  \bigintsss_{\,v_3}^{v_2}\!\! P\;\partial\,v
  \; = \; 
  P_1 \:.\: (\,v_2\:-\:v_3\,)
  \;\: ,  \\
 \mbox{and if we add (\ref{Eq_66}):} &
 \hspace*{5mm} 
  P_1 \;=\; P_2 \;=\; P_3
  \;\: , 
\end{aligned}
\;\;\;\;\;\;\;\;
\right\} 
\label{Eq_69} 
\end{equation}
where $P$ is to be regarded as a certain function of $v$ and $T$, 
and if we imagine the
{\color{red}{\it(common)}}
temperature $T_1=T_2=T_3$ to be denoted by
$T$, we have here $4$ equations
{\color{red}{\it(with two equations contained in
                 $P_1\,=\,P_2\,=\,P_3$)}} 
with the 4 unknowns $T$, $v_1$, $v_2$, $v_3$,
for which the subject is to look for the maximum states. 

It is noteworthy that the constants included in these $4$ equations depend only on the chemical nature of the body, but not on the given values of mass $M$, volume $V$ and energy $U$. We can therefore call these equations the internal equilibrium conditions, in contrast to equations (\ref{Eq_56}), (\ref{Eq_57}) and (\ref{Eq_58}), which denote the external conditions to which the body is subjected.

Before we proceed to the consideration and comparison of the values of the unknowns resulting from these equations, we want to examine in general whether they really characterise a maximum state of the body, i.e. whether the state corresponding to them does not, for example, cause a minimum of entropy. To answer this question, we must analyse the value of $\delta^{2}S$. If it is negative for all possible variations, then the state in question is in any case a maximum state.

We therefore vary equation (\ref{Eq_64}) and thereby obtain the expression of $\delta^{2}S$, which is significantly simplified if we apply equations (\ref{Eq_65}), (\ref{Eq_66}), (\ref{Eq_67}) and (\ref{Eq_68}). If we then consider the values of $\delta\,s_1$, $\delta\,s_2$, $\delta\,s_3$ by means of (\ref{Eq_59}) and eliminate $\delta\,M_1$ and $\delta\,M_3$ by means of (\ref{Eq_61}), (\ref{Eq_62}) and (\ref{Eq_63}), the result is: 
\vspace*{0mm}
\begin{align}
 \delta^{2}S & \:=\; 
  -\:\sum_k 
   \left(
     \frac{M_k\:.\:\delta\,s_k\:.\:\delta\,T_k}{T_k}
   \right)
\;+\;\sum_k 
   \left(
     \frac{M_k\:.\:\delta\,P_k\:.\:\delta\,v_k}{T_k}
   \right)
\;\: , \nonumber 
\end{align}
for which we can also write: 
\vspace*{0mm}
\begin{align}
 T\:.\:\delta^{2}S & \:=\; 
  -\:\sum_k 
    \: M_k \:.\:
   \left(\:
     \delta\,s_k\:.\:\delta\,T_k
     \;-\;
     \delta\,P_k\:.\:\delta\,v_k
   \:\right)
\;\: . \nonumber 
\end{align}
To reduce all variations to those of the independent variables $T$ and $V$, we set according to (\ref{Eq_53}): 
\vspace*{0mm}
\begin{align}
 \delta\,s & \:=\;
  \frac{k}{T} \:\:.\:\: \delta\,T
  \;+\;
  \frac{\partial P}{\partial\,T} \:\:.\:\: \delta\,v
\; , \nonumber 
\end{align}
where $k= {\partial\,u}/{\partial\,T}$ denotes the specific heat at constant volume according to (\ref{Eq_50}), and: 
\vspace*{0mm}
\begin{align}
 \delta\,P & \:=\;
  \frac{\partial P}{\partial\,T} \:\:.\:\: \delta\,T
  \;+\;
  \frac{\partial P}{\partial\,v} \:\:.\:\: \delta\,v
\; . \nonumber 
\end{align}

Then we obtain:
\vspace*{0mm}
\begin{align}
 T\:.\:\delta^{2}S & \:=\; 
  -\:\sum_k 
    \: M_k \:.
   \left[\:\:
      \frac{k_k}{T} 
     \:\,.\:\: 
     \delta\,T_k^2
     \:\;-\;
     \left(
     \frac{\partial P}{\partial\,v}
     \right)_k
     .\:\: 
     \delta\,v_k^2
   \:\:\right]
\;\: . \label{Eq_70} 
\end{align}
If the variables $k_1$, $k_2$, $k_3$ are all positive, and the variables 
$({\partial\,P}/{\partial\,v})_1$, 
$({\partial\,P}/{\partial\,v})_2$, 
$({\partial\,P}/{\partial\,v})_3$ 
are all negative, 
then $\delta^{2}S$, as can be seen from the last equation, is essentially negative, i.e. the state under consideration is a maximum state, i.e. a (stable or unstable) 
state of equilibrium. 
Since $k$ is always positive according to its physical meaning, the condition of equilibrium depends on whether 
$({\partial\,P}/{\partial\,v})$
is negative for all three parts of the body or not, and in the latter case the state is not a maximum state.

In fact, experience shows that $({\partial\,P}/{\partial\,v})$ is negative in every equilibrium state, since the pressure, whether positive or negative, always changes in the opposite direction to the volume at constant temperature. 
However (as we will see below), there are also states in which $({\partial\,P}/{\partial\,v})$ is positive, and these states therefore never represent an equilibrium position, and are therefore not accessible to direct observation.
If, on the other hand, $({\partial\,P}/{\partial\,v})$ is negative, then equilibrium takes place, but this does not necessarily mean that it is stable, because it then depends on whether another maximum state exists under the given conditions, to which a greater value of entropy corresponds.

\vspace*{2mm}
\begin{center}
--------------------------------------------------- 
\end{center}
\vspace*{-2mm}

We now move on to examining the values of the variables that represent a solution to equations (\ref{Eq_69}). This is possible in \dashuline{\,three different ways}.

{\bf 1) If we set:} \vspace*{-2mm}
           $$ v_1 \;=\; v_2 \;=\; v_3 $$ 
all $4$ equations (\ref{Eq_69}) are thereby satisfied, because since the temperature $T$ is in any case common to all $3$ parts of the body, their states are thereby completely identical, i.e. the whole body is homogeneous. 
Its state is then determined if one also includes equations (\ref{Eq_56}), (\ref{Eq_57}) and (\ref{Eq_58}), which express the external conditions.
In this case they become 
\vspace*{-3mm}
\begin{align}
  M_1 \;+\; M_2 \;+\; M_3 \;\;\:& \:=\; M
  \;\: , \nonumber 
 \\
  v_1\; \left(\:M_1 \;+\; M_2 \;+\; M_3\:\right) & \:=\; V
  \;\: , \nonumber 
 \\
  u_1\; \left(\:M_1 \;+\; M_2 \;+\; M_3\:\right) & \:=\; U
  \;\: , \nonumber 
 \\
 \mbox{and consequently}\;\;\;
  v_1 \;=\; \frac{V}{M}
  \;\;\;\;\mbox{and}\;\;
  u_1 & \:=\; \frac{U}{M}
  \;\: . \nonumber 
\end{align}
This also results in the value of the temperature $T$ if you consider $u$ as a function of $T$ and $v$.

{\bf 2) If we set:}   \vspace*{-2mm}
           $$ v_2 \;=\; v_3 $$ 
while assuming $v_1$ to be different, the states of matter labelled $2$ and $3$ coincide and equations (\ref{Eq_69}) reduce to: 
\vspace*{-3mm}
\begin{align}
  P_1 & \:=\; P_2 
  \;\: , \label{Eq_71} 
 \\
  \bigintsss_{\,v_2}^{v_1}\!\! P\;\partial\,v \:
  & \:=\; P_1 \:.\: \left(\: v_1 \:-\: v_2 \:\right) 
  \;\: . \label{Eq_72} 
\end{align}
In this case, the body is in two different states of aggregation next to each other, e.g. in the form of vapour and liquid.

(Considering the second equation (\ref{Eq_72}), it follows that the pressure $P$, since it has the same value $P_1=P_2$ for the two limits of the integral, must assume values between them which are partly $<P_1$, partly $>P_1$, so that equation (\ref{Eq_72}) can be satisfied. 
Therefore, if the latter is to be the case, $({\partial\,P}/{\partial\,v})$ between these limits must be partly positive and partly negative).

The two equations (\ref{Eq_71}) and (\ref{Eq_72}) contain the $3$ unknowns ($T$, $v_1$, $v_2$), so they can be used to represent the quantities $v_1$ and $v_2$ (and consequently also the pressure $P_1=P_2$ and the specific energies $u_1$ and $u_2$) as certain functions of $T$.

The state of the whole body is completely determined if one also takes into account the other conditions (\ref{Eq_56}), (\ref{Eq_57}), (\ref{Eq_58}), which for this case are: 
\vspace*{0mm}
\begin{align}
  M_1 \;+\; \left(\:M_2 \;+\; M_3\:\right) 
  \;\;\;\;\;\;\;\;\:& \:=\; M
  \;\: , \nonumber 
 \\
  M_1\;v_1 \;+\; \left(\:M_2 \;+\; M_3\:\right)\:.\:v_2 
  \;\;\:& \:=\; V
  \;\: , \nonumber 
 \\
  M_1\;u_1 \;+\; \left(\:M_2 \;+\; M_3\:\right)\:.\:u_2 
  \;\;\:& \:=\; U
  \;\: . \nonumber 
\end{align}

These $3$ equations are used to calculate the last $3$ unknowns, namely $T$, $M_1$ and $(M_2+M_3)$.  This means that the state of the body is completely given, because the quantities $M_2$ and $M_3$ obviously only depend on their sum. Of course, the result only makes physical sense if $M_1$ and $(M_2+M_3)$ are positive.

{\bf 3) If we let $v_1$, $v_2$ and $v_3$ be different from each other}, we have from (\ref{Eq_69}) the equations: 
\vspace*{0mm}
\begin{align}
  P_1 & \:=\; P_2 \:=\; P_3 
  \;\: , \nonumber 
 \\
  \bigintsss_{\,v_2}^{v_1}\!\! P\;\partial\,v \:
  & \:=\; P_1 \:.\: \left(\: v_1 \:-\: v_2 \:\right) 
  \;\: , \nonumber 
 \\
  \bigintsss_{\,v_3}^{v_2}\!\! P\;\partial\,v \:
  & \:=\; P_1 \:.\: \left(\: v_2 \:-\: v_3 \:\right) 
  \;\: . \nonumber 
\end{align}

This case describes a state in which all $3$ states of aggregation exist side by side in the body. The $4$ equations contain $4$ unknowns, namely $T$, $v_1$, $v_2$, $v_3$, so that 
a unique 
solution corresponds to them. Once this has been found, the masses $M_1$, $M_2$, $M_3$ 
of the body parts in the different aggregate states are calculated 
from the external conditions (\ref{Eq_56}), (\ref{Eq_57}), (\ref{Eq_58}): 
\vspace*{0mm}
\begin{align}
  \sum_k \: M_k  
  \;\;\;& \:=\; M
  \;\: , \nonumber 
 \\
  \sum_k \: M_k  \:.\: v_k 
  & \:=\; V
  \;\: , \nonumber 
 \\
  \sum_k \: M_k  \:.\: u_k 
  & \:=\; U
  \;\: . \nonumber 
\end{align}
However, this solution only makes sense if the values of $M_1$, $M_2$, $M_3$ are positive.
\vspace*{0mm}
\begin{center}
--------------------------------------------------- 
\end{center}
\vspace*{-2mm}

We shall now proceed to examine the physical significance of these results and compare them with the results of experience. 

{\bf \underline{1st Solution}.} All 3 parts of the body similar and: 
\vspace*{-1mm}
\begin{align}
  v_1 \;=\; v_2 \;=\; v_3 \;\;\:& \:=\; \frac{V}{M}
  \;\: , \nonumber 
 \\
  u_1 \;=\; u_2 \;=\; u_3 \;\;\:& \:=\; \frac{U}{M}
  \;\: . \nonumber 
\end{align}
From this {\it\color{red}we can deduce} also the value of T.

This solution corresponds to a state in which the whole body is homogeneous.
It always has a certain meaning, but, as we have seen from equation (\ref{Eq_70}), it represents a state of equilibrium only in the case where $({\partial\,P}/{\partial\,v})$ is negative, as is confirmed by experience. If this condition is fulfilled, the equilibrium is unstable or stable, depending on whether a state exists under the given external conditions which corresponds to an even greater value of entropy or not. When the former is the case will be shown below.

{\bf \underline{2nd Solution}.} 
The 2nd and 3rd parts of the body are similar, with the internal conditions (\ref{Eq_71})-(\ref{Eq_72}): 
\vspace*{-3mm}
\begin{align}
  \hspace*{55mm}
  P_1 & \:=\; P_2 
  \;\: , 
  \hspace*{66mm} (71)
  \nonumber 
 \\
  \hspace*{55mm}
  \bigintsss_{\,v_2}^{v_1}\!\! P\;\partial\,v \:
  & \:=\; P_1 \:.\: \left(\: v_1 \:-\: v_2 \:\right) 
  \;\: . 
  \hspace*{45mm} (72)
  \nonumber 
\end{align}
From this, $v_1$ and $v_2$ (thus also $P_1$, $u_1$ and $u_2$) are calculated as functions of $T$ alone. 

The external conditions are {\it\color{red}(as stated before)}\,:
\begin{equation}
\left.
\begin{aligned}
  M_1 \;+\; \left(\:M_2 \;+\; M_3\:\right) 
  \;\;\;\;\;\;\;\;\:& \:=\; M
  \;\: ,  \\
  M_1\;v_1 \;+\; \left(\:M_2 \;+\; M_3\:\right)\:.\:v_2 
  \;\;\:& \:=\; V
  \;\: ,  \\
 M_1\;u_1 \;+\; \left(\:M_2 \;+\; M_3\:\right)\:.\:u_2 
  \;\;\:& \:=\; U
  \;\: . 
\end{aligned}
\;\;\;\;\;\;\;\;
\right\} 
\label{Eq_73} 
\end{equation}
From this {\it\color{red}we can deduce} the values of $T$, $M_1$ and $(M_2+M_3)$.

This solution corresponds to a state in which different states of aggregation occur side by side in the body $2$, but, as already noted, it only makes sense if the values of $M_1$ and $(M_2+M_3)$ are positive. 
Furthermore, according to (\ref{Eq_70}), it only corresponds to a state of equilibrium if 
$({\partial\,P}/{\partial\,v})_1$ and 
$({\partial\,P}/{\partial\,v})_2$ are negative.
It can now be proved that, assuming these conditions, the entropy of the state corresponding to this solution is in any case greater than the entropy of the state corresponding to the previous solution, with the same external conditions, in other words that the equilibrium state we are considering here, assuming that it has any real meaning at all, is in all circumstances more stable than that of the previous solution, as is also the case in experience.
For this purpose, \dashuline{\,it is only necessary to prove that the difference between the entropies corresponding to the two solutions is always positive\,}. However, since this investigation is of a purely mathematical nature and does not present any major difficulties, \dashuline{\,we will skip it here and turn to the discussion of equations (\ref{Eq_71}) and (\ref{Eq_72})\,}, which express the internal equilibrium conditions of a body that is in two different states of aggregation side by side. 

These equations apply to vapours and liquids in contact with each other, as well as to solid bodies in contact with the corresponding liquid or vapour.
\vspace*{0mm}
\begin{center}
--------------------------------------------------- 
\end{center}
\vspace*{-2mm}

As we have already noted, all the variables occurring in (\ref{Eq_71}) and (\ref{Eq_72}) can be represented as functions of temperature alone. Here $v_1$ and $v_2$ denote the specific volumes of the two different parts of the body in contact, e.g. vapour and liquid, corresponding to the temperature $T$. It does not matter which index is related to one or the other state of matter, since the equations are completely symmetrical with respect to both indices.

For each value of the temperature, the two equations give $3$ different solutions, possibly even imaginary ones, which correspond to the $3$ pairwise combinations of the $3$ states of aggregation. However, in order to fix the ideas, we shall first consider, for example, the solution which corresponds to the contact of vapour and liquid.
Since it governs the equilibrium between these two aggregate forms, it contains the laws of saturated vapours. If we refer the index $1$ to the vapour and the index $2$ to the liquid, then $v_1$ is the specific volume of the vapour saturated at the temperature $T$, $P_1$ is its pressure and $v_2$ is the specific volume of the liquid in contact with it. These $3$ quantities are therefore functions of the temperature alone, as experience shows.

We can first arrive at new relations by differentiating the equations under consideration, whereby, since all variables depend only on $T$, we want to designate the differential coefficients briefly as $dv_1/dT$, $dv_2/dT$, $dP_1/dT$, while we retain the designation ${\partial\,P}/{\partial\,T}$ for the partial differential quotient of $P$ with respect to $T$ (at constant volume).

Then the equations (\ref{Eq_71}) and (\ref{Eq_72}) totally differentiated according to $T$ give: 
\vspace*{0mm}
\begin{align}
  \frac{dP_1}{dT} & \:=\; \frac{dP_2}{dT} 
  \;\: , 
  \label{Eq_74} 
\end{align}
and:
\vspace*{-1mm}
\begin{align}
  \bigintsss_{\,v_2}^{v_1}\!\!
     \frac{\partial P}{\partial T} \;\partial\,v 
  \;+\;
    P_1 \:.\: \frac{\partial v_1}{\partial T}
  \;-\;
    P_2 \:.\: \frac{\partial v_2}{\partial T}
  & \:=\;
    P_1 \:. 
    \left(\:
    \frac{\partial v_1}{\partial T}
    \:-\:
    \frac{\partial v_2}{\partial T}
    \:\right)
  \;+\;
    \frac{\partial P_1}{\partial T} \:.\: 
    \left(\:v_1 \:-\:v_2\:\right)
  \;\: , 
  \nonumber 
\end{align}
or, since according to (\ref{Eq_71}) $P_1=P_2$: 
\vspace*{0mm}
\begin{align}
  \bigintsss_{\,v_2}^{v_1}\!\!
     \frac{\partial P}{\partial T} \;\partial\,v 
  & \:=\;
    \frac{\partial P_1}{\partial T} \:.\: 
    \left(\:v_1 \:-\:v_2\:\right)
  \;\: . 
  \label{Eq_75} 
\end{align}
From this we can immediately calculate the difference in the specific energies $(u_1-u_2)$ of vapour and liquid. 
Since according to (\ref{Eq_51}) in general: 
\vspace*{0mm}
\begin{align}
    \frac{\partial u}{\partial v} 
  & \:=\;
    T \:.\: \frac{\partial P}{\partial T} \;-\; P
  \;\: 
  \nonumber 
\end{align}
we have, if we integrate from the liquid state to the vapour state partially according to $v$: 
\vspace*{0mm}
\begin{align}
 u_1 \;-\; u_2
  & \:=\;
   T \bigintsss_{\,v_2}^{v_1}\!
     \frac{\partial P}{\partial T} \;\partial\,v 
    \:\;-
   \bigintsss_{\,v_2}^{v_1}\!\! P \;\partial\,v 
  \;\: . 
  \nonumber 
\end{align}
or according to (\ref{Eq_75}) and (\ref{Eq_72}): 
\vspace*{0mm}
\begin{align}
 u_1 \;-\; u_2
  & \:=\;   
   \left(\,  
      T \:\: \frac{\partial P_1}{\partial T} \:-\: P_1
   \,\right)
   \;
   \left(\:  
      v_1 \;-\; v_2
   \:\right)
  \;\: . 
  \label{Eq_76} 
\end{align}

Let us now introduce ``\,$r$,'' the heat of vaporisation (at constant pressure) of the unit of mass. 
This includes:
1) the increase in energy $(u_1-u_2)$ required for vaporisation; and 
2) the external work $P_1\:(v_1-v_2)$ to be performed, since vaporisation takes place at constant external pressure. 
Therefore, we have for the heat of vaporisation: 
\vspace*{-3mm}
\begin{align}
  r & \:=\;   
  (\,u_1 \;-\; u_2\,)
  \;+\; 
  P_1 \; (\,v_1 \;-\; v_2\,)
  \;\: , 
  \label{Eq_77} 
\end{align}
i.e., according to (\ref{Eq_76}):
\vspace*{-3mm}
\begin{align}
  r & \:=\; 
  T \:\: \frac{\partial P_1}{\partial T} 
  \;
  (\,v_1 \;-\; v_2\,)
  \;\: ,
  \label{Eq_78} 
\end{align}
as is known.

The specific heat ``$h_1$'' of the saturated vapour, as first introduced by Clausius, is obviously obtained from the equation : 
\vspace*{-3mm}
\begin{align}
  h_1\:d\,T & \:=\; 
  du_1 \;+\; P_1\;dv_1 
  \;\: ,
  \nonumber 
\end{align}
because $h_1\:d\,T$ is the heat that must be supplied to the unit mass of saturated vapour from the outside so that it is just saturated again at the temperature $T + d\,T$.

This results in: 
\vspace*{-3mm}
\begin{align}
  h_1& \:=\; 
  \frac{du_1}{dT} \;+\; P_1\;\frac{dv_1}{dT} 
  \;\: ,
  \nonumber 
\end{align}
and in the same way for the liquid: 
\vspace*{-3mm}
\begin{align}
  h_2& \:=\; 
  \frac{du_2}{dT} \;+\; P_2\;\frac{dv_2}{dT} 
  \;\: .
  \nonumber 
\end{align}
The quantity $h_2$ is usually imperceptibly different from the specific heat of the liquid in question at constant pressure, because in liquids a pressure that is not very significant has no noticeable influence on the state.

From equation (\ref{Eq_77}), the total differentiation according to $T$ now yields: 
\vspace*{0mm}
\begin{align}
  \frac{d\,r}{d\,T} & \:=\; 
  \frac{d\,u_1}{d\,T} \;-\; \frac{d\,u_2}{d\,T} 
  \;+\; P_1\,
  \left(\:
  \frac{d\,v_1}{d\,T} \;-\; \frac{d\,v_2}{d\,T} 
  \:\right)
  \;+\; \frac{d\,P_1}{d\,T}\,
  \left(\:
  v_1 \;-\; v_2
  \:\right)
  \;\: ,
  \nonumber 
\end{align}
and consequently, with the introduction of $h_1$ and $h_2$: 
\vspace*{0mm}
\begin{align}
  \frac{d\,r}{d\,T} & \:=\; 
  h_1 \;-\; h_2
  \;+\; \frac{d\,P_1}{d\,T}\,
  \left(\:
  v_1 \;-\; v_2
  \:\right)
  \;\: ,
  \nonumber 
\end{align}
or according to (\ref{Eq_78}): 
\vspace*{0mm}
\begin{align}
  h_1 \;-\; h_2 & \:=\;  
  \frac{d\,r}{d\,T}
  \;-\; 
  \frac{r}{T}
  \;\: .
  \nonumber 
\end{align}
From this, as is known, $h_1$ can be calculated by 
preprocessing 
the specific heat of the liquid at constant pressure for $h_2$.

The equations derived so far are identical to those already known from the existing theory of saturated vapours and the evaporation process.
However, we now want to develop a formula that has not yet become known, but which, in addition to its theoretical interest, offers a practical benefit in that it allows the specific heat of vapours at constant pressure (and also that at constant volume), which is so difficult to access by direct observation, to be calculated in a simple way. We obtain this quantity directly by applying the present equations.

If, for the sake of convenience, the quantity $d\,P_1/d\,T$ in equation (\ref{Eq_75}) is expressed with $r$ using (\ref{Eq_78}), we obtain:  
\vspace*{-3mm}
\begin{align}
 \bigintsss_{\,v_2}^{v_1}\!
     \frac{\partial P}{\partial T} \;\partial\,v 
  & \:=\;
  \frac{r}{T}   
  \;\: , 
  \nonumber 
\end{align}
and by total differentiation of this equation with respect to  
$T$:$\,$\footnote{\color{red}\label{footnote_error}$\:${\it Note that the last term of (\ref{Eq_79}) was wrongly written as $-\:r/T_2$ (instead of $-\:r/T^2$) in the paper of Planck (P. Marquet)}.}
\vspace*{0mm}
\begin{align}
 \bigintsss_{\,v_2}^{v_1}\!
     \frac{\partial^2 P}{\partial\,T^2} \;\partial\,v 
  \;+\;
  \left(\frac{\partial\,P}{\partial\,T}\right)_{\!1}
  .\;\frac{d\,v_1}{d\,T}
  \;-\;
  \left(\frac{\partial\,P}{\partial\,T}\right)_{\!2}
  .\;\frac{d\,v_2}{d\,T}
  & \:=\; 
  \frac{1}{T} \:.\:
  \frac{d\,r}{d\,T} 
  \;-\; 
  \frac{r}{T^2}   
  \;\: .
   \label{Eq_79} 
\end{align}
Now, according to (\ref{Eq_52}): 
\vspace*{-3mm}
\begin{align}
  \frac{\partial\,k}{\partial\,v}
  & \:=\; T \:.\:\: 
     \frac{\partial^2 P}{\partial\,T^2} 
  \;\: ,
   \nonumber 
\end{align}
where $k$ denotes the specific heat at constant volume. Consequently, if you partially integrate from the liquid to the vapour state according to $v$: 
\vspace*{-2mm}
\begin{align}
  k_1 \;-\; k_2 
  & \:=\; T \:.\:\: 
     \bigintsss_{\,v_2}^{v_1}\!
     \frac{\partial^2 P}{\partial\,T^2} \;\partial\,v 
  \;\: .
   \nonumber 
\end{align}

The integral from (\ref{Eq_79}) replaced by its value gives: 
\vspace*{0mm}
\begin{align}
         k_1 \;-\; k_2 
  & \:=\;  
        \frac{d\,r}{d\,T} 
  \;-\; \frac{r}{T}   
  \;-\;  T \:. 
  \left(\frac{\partial\,P}{\partial\,T}\right)_{\!1}
  .\;\frac{d\,v_1}{d\,T}
  \;+\;  T \:. 
  \left(\frac{\partial\,P}{\partial\,T}\right)_{\!2}
  .\;\frac{d\,v_2}{d\,T}
  \;\: .
  \nonumber 
\end{align}
Here we can write according to (\ref{Eq_49}): 
\vspace*{-3mm}
\begin{align}
 \frac{\partial P}{\partial\,T} 
   & \:=\; 
 -\:\alpha \:.\:
 \frac{\partial P}{\partial v} 
\; , \nonumber 
\end{align}
where $\alpha$ denotes the differential quotient of $v$ with respect to $T$ at constant pressure, see (\ref{Eq_33}). 
This implies: 
\vspace*{0mm}
\begin{align}
         k_1 \;-\; k_2 
  & \:=\;  
        \frac{d\,r}{d\,T} 
  \;-\; \frac{r}{T}   
  \;+\;  {\alpha}_1 \:.\: T \:. 
  \left(\frac{\partial\,P}{\partial\,v}\right)_{\!1}
  .\;\frac{d\,v_1}{d\,T}
  \;-\;  {\alpha}_2 \:.\: T \:. 
  \left(\frac{\partial\,P}{\partial\,v}\right)_{\!2}
  .\;\frac{d\,v_2}{d\,T}
  \;\: .
  \nonumber 
\end{align}
For the specific heat at constant pressure $c$, we finally have according to (\ref{Eq_55}): 
\vspace*{0mm}
\begin{align}
 c \;-\; k 
 & \:=\; 
 \alpha \;\: T \;.\;\: \frac{\partial P}{\partial\,T} 
\; , \nonumber 
\end{align}
and if we use this equation to introduce $c_1$ and $c_2$ instead of $k_1$ and $k_2$, we get : 
\vspace*{0mm}
\begin{align}
         c_1 \:-\: c_2 
  & \:=\;  
        \frac{d\,r}{d\,T} 
  \;-\; \frac{r}{T}   
  \;+\:  {\alpha}_1 \:.\: T \:. 
  \left\{
  \left(\frac{\partial\,P}{\partial\,T}\right)_{\!1}
  \:+\:
  \left(\frac{\partial\,P}{\partial\,v}\right)_{\!1}
  .\;\frac{d\,v_1}{d\,T}
  \right\}
  \;-\:  {\alpha}_2 \:.\: T \:. 
  \left\{
  \left(\frac{\partial\,P}{\partial\,T}\right)_{\!2}
  \:+\:
  \left(\frac{\partial\,P}{\partial\,v}\right)_{\!2}
  .\;\frac{d\,v_2}{d\,T}
  \right\}
  \; .
  \nonumber 
\end{align}
Since the expressions in the two brackets represent the same quantity $dP_1/d\,T = dP_2/d\,T$, see (\ref{Eq_74}), we can write: 
\vspace*{-3mm}
\begin{align}
         c_1 \:-\: c_2 
  & \:=\;  
        \frac{d\,r}{d\,T} 
  \;-\; \frac{r}{T}   
  \;+\: T \:.\: \frac{d\,P_1}{d\,T} \:.\,
        \left(\,{\alpha}_1\:-\:{\alpha}_2\,\right)
  \; ,
  \nonumber 
\end{align}
or according to (\ref{Eq_78}): 
\vspace*{-4mm}
\begin{align}
         c_1 \:-\: c_2 
  & \:=\;  
        \frac{d\,r}{d\,T} 
  \;-\; \frac{r}{T}   
  \;+\: 
        \frac{r}{v_1 \:-\: v_2} \:.\,
        \left(\,{\alpha}_1\:-\:{\alpha}_2\,\right)
  \; .
  \nonumber 
\end{align}
Finally, if for the sake of clarity we use the symbolic notation $[\,\partial\,v/\partial\,T\,]_P$ for $\alpha$, where the index $P$ means that the pressure should remain constant during the differentiation, we obtain:
\vspace*{0mm}
\begin{align}
         c_1 \:-\: c_2 
  & \:=\;  
        \frac{d\,r}{d\,T} 
  \;-\; \frac{r}{T}   
  \;+\: 
        \frac{r}{v_1 \:-\: v_2} \:.\,
  \left\{\;
  \left[\,\frac{\partial\,v_1}{\partial\,T}\,\right]_{\!P}
  \:-\:
  \left[\,\frac{\partial\,v_2}{\partial\,T}\,\right]_{\!P}
  \;\right\}
  \; .
  \label{Eq_80} 
\end{align}
In this equation, all quantities are functions of the temperature;.
Therefore, one of them can be represented in its entire dependence on the temperature if all the others are known as functions of the temperature. For the case we are considering here, namely for the combination of liquid and vapour, the equation is best suited to calculating the specific heat $c_1$ for a vapour (which is in a state of saturation) at constant pressure, since for this purpose it is only necessary to know the specific heat $c_2$ of the corresponding liquid (at the same temperature and pressure of the saturated vapour), the heat of vaporisation $r$ and the specific volumes $v_1$ and $v_2$ of the saturated vapour and the liquid, together with their variation with temperature at constant pressure.
\vspace*{-2mm}
\begin{center}
--------------------------------------------------- 
\end{center}
\vspace*{-2mm}

Since \dashuline{\,the above formula (\ref{Eq_80}) has not yet been published\,}, it would seem appropriate at this point to apply it to a specific case in which it is possible to check its consistency with experience.

For example, let us calculate the \dashuline{\,specific heat of water vapour\,} {\color{red}\it --\,i.e. $c_1$ from (\ref{Eq_80})\,--} when it is in a state of saturation at $100\,{}^{\circ}$~C (i.e. under the pressure of one atmosphere).

Since each element of equation (\ref{Eq_80}) contains a quantity of heat as a factor, it is obviously permissible to measure all the quantities of heat occurring in it according to caloric measure.
Then we have for the individual quantities: 
\begin{enumerate}[label=$\bullet$~,
  leftmargin=10mm,parsep=0mm,itemsep=1mm,
  topsep=0mm,rightmargin=2mm]
\item 
$T = 273 + 100 = 373$ 
 \hspace*{31mm}
(absolute temperature)
\item 
 $c_2= 1.0130$ according to Regnault
 \hspace*{9mm}
 (specific heat of water at $100$°C.)
\item 
$r=536.2$ according to Clausius 
 \hspace*{14mm}
(heat of vaporisation of water at $100$°C)
\item 
 $dr/dT=-0.708$ according to Clausius
\item 
$v_1 = 1650.4$ according to Hirn 
 \hspace*{17mm}
(Volume of one gram of saturated water vapour 
 \\ \hspace*{70mm} at $100$°C, in cubic centimetres) 
\item 
$[\,\partial\,v_1/\partial\,T\,]_P= 4.843$ according to Hirn 
 \hspace*{2mm}
(Growth of the specific volume when the vapour is
 \\ \hspace*{50mm} superheated by $1$°C under the constant pressure of an atmosphere).
\item 
$v_2 = 1.0431$ according to Jolly
 \hspace*{16mm}
(Volume of one gram of aqueous vapour 
 \\ \hspace*{70mm} at $100$°C. in cubic centimetres).
\item 
$[\,\partial\,v_2/\partial\,T\,]_P = 0.00073$ according to Joly 
(Volume of the specific volume 
 \\ \hspace*{70mm} when water is heated by $1$°C.)
\end{enumerate}

The references of these numerical values are: 
Regnault$\,$\footnote{$\:$Regnault / M\'emoires de l'Acad. T. XXI.},
Clausius$\,$\footnote{$\:$Clausius / Pogg. Ann. Bd. XCVII.},
Hirn$\,$\footnote{$\:$Hirn / Th\'eorie m\'ecaniqne de la chaleur. Paris, 1865.} and 
Jolly$\,$\footnote{$\:$Jolly / Monatsber. d. M\"unchner Akad.
1864.}.

Using these values, equation (\ref{Eq_80}) gives the specific heat of water vapour at $100$°C under the constant pressure of an atmosphere: 
\vspace*{-3mm}
$$c_1 \;=\; 0.442 \; .$$ 

Only Regnault's direct observation
results$\,$\footnote{$\:$Regnault / M\'emoires de l'Acad. T. XXVI.}  
are available, which represent the mean value of the specific heat of water vapour in the range of temperatures from about 130°C to 220°C at the constant pressure of an atmosphere. The 4 Regnault observation series give the mean of the values found: 
\vspace*{-1mm}
$$c_1 \;=\; 0.478 \; .$$ 
The reason for the deviation of this result from our own is adequately explained by the fact that the specific heat of vapours increases somewhat with temperature, a fact which has been established as probable by all previous observations.

The behaviour of the specific heat of a vapour at constant pressure at different temperatures can now be checked exactly according to equation (\ref{Eq_80}), but it must be noted that this formula is only strictly valid when the vapour is in a state of saturation.  

  We will ignore further applications here and only provide an approximation formula, 
  which in most cases gives fairly accurate results, 
namely when the specific volume of the liquid against that of the vapour can be neglected. Under this condition, the equation (\ref{Eq_80}) is: 
\vspace*{0mm}
\begin{align}
         c_1 \:-\: c_2 
  & \:=\;  
        \frac{d\,r}{d\,T} 
  \;-\; \frac{r}{T}   
  \;+\: 
        \frac{r}{v_1} \:.\,
  \left[\,\frac{\partial\,v_1}{\partial\,T}\,\right]_{\!P}
  \; .
  \label{Eq_81} 
\end{align}

This formula provides a rough approximation.
A significant simplification, assuming that Gay-Lussac's law applies to the vapour up to the point of saturation, leads to the equation: 
\vspace*{0mm}
\begin{align}
  \left[\,\frac{\partial\,v_1}{\partial\,T}\,\right]_{\!P}
  & \:=\;  
  \frac{v_1}{T}   
  \; ,
  \nonumber 
\end{align}
and thus the equation (\ref{Eq_81}) becomes: 
\vspace*{-3mm}
\begin{align}
         c_1 \:-\: c_2 
  & \:=\;  
        \frac{d\,r}{d\,T} 
  \; .
  \nonumber 
\end{align}
For the example above for which $c_2 = 1.013$ and $dr/dT = - 0.708$, {\it\color{red}\,this simplified formulation\,} results in 
\vspace*{-1mm}
$$c_1 \;=\; 0.305 \; ,$$ 
which is significantly too small.
\vspace*{2mm}
\begin{center}
--------------------------------------------------- 
\end{center}
\vspace*{-2mm}

Let us now consider the application of equation (80) to the combination of a liquid body with the corresponding solid body, applying index $1$ to the liquid body and index $2$ to the solid body. The equation then applies to the melting temperature $T$, which, like the boiling temperature, varies with the pressure. Furthermore, $c_1$ and $c_2$ are the specific heats of the liquid and solid body at constant pressure, as it corresponds to the melting temperature, $r$ is the heat of fusion, $v_1$ and $v_2$ are the specific volumes of the liquid and solid body in the state of melting

Let us also make a special application to this case, e.g. for phosphorus at 44°C, i.e. the melting temperature at the pressure of one atmosphere.
\vspace*{2mm}

Since the specific heats of the liquid and solid phosphorus are known, equation (\ref{Eq_80}) can be used to calculate the value of $dr/dT$, i.e. the change in the heat of fusion of the phosphorus with the melting temperature.

\begin{enumerate}[label=$\bullet$~,
  leftmargin=10mm,parsep=0mm,itemsep=1mm,
  topsep=0mm,rightmargin=2mm]
\item 
$T = 273 + 44 = 317$ 
 \hspace*{40mm}
(absolute temperature)
\item 
 $c_1= 0.2045$ according to Person
 \hspace*{20mm}
 (specific heat of liquid phosphorus at $48$°C)
\item 
 $c_2= 0.1887$ according to Regnault
 \hspace*{16mm}
 (specific heat of solid phosphorus at $20$°C)
\item 
$r=5.034$ according to Person 
 \hspace*{23mm}
(heat of fusion of phosphorus at $44$°C)
\item 
$v_1 = 1.051\,73$ according to Kopp 
 \hspace*{19mm}
(specific volume of liquid phosphorus at $44$°C, \\
 \hspace*{64mm}
the specific volume of solid phosphorus at $0$°C$\:=1$)
\item 
$[\,\partial\,v_1/\partial\,T\,]_P= 0.000\,532$ according to Kopp
 \hspace*{1mm} (Expansion when heated by $1$°C)
\item 
$v_2 = 1.016\,85$ according to Kopp
 \hspace*{19mm}
(specific volume of solid phosphorus at $44$°C)
\item 
$[\,\partial\,v_2/\partial\,T\,]_P = 0,000\,383$ according to Kopp 
 \hspace*{0mm}
(Expansion when heated by $1$°C)
\end{enumerate}

The references of these numerical values are: 
Person$\,$\footnote{$\:$Person / Annales de chim. et de phys. III. S\'erie. T. XXI.},
Kopp$\,$\footnote{$\:$Kopp / Liebigs Annal. Bd. XCIII.} and 
Regnault$\,$\footnote{$\:$Regnault / Annales de chim. et de phys. III. S\'erie. T. XXVI.}.

Using these values, equation (\ref{Eq_80}) can be used to calculate: 
$$ \frac{dr}{dT} \;=\; 0.0102 \;, $$ 
i.e. if the melting temperature of the phosphorus is increased (by external pressure), the heat of fusion also increases, namely (assuming proportionality) for $1$°C by $0.01$ heat units, i.e. by $0.2\,$\%. It should be noted here that $r$ is the amount of heat that must be supplied from outside so that the unit mass of the body melts under the constant pressure that corresponds to the melting temperature.

No experimental data are available to verify this result because, as far as I know, no experiments have yet been carried out on the variability of the heat of fusion with the melting temperature. 

It would now remain to apply the formula to the combination of a solid body with its vapour. But here the experiments are too few to make such an application possible, for the reason that in the rarest cases such a combination represents a stable state of equilibrium accessible to observation.
It is, however, remarkable that all the equations derived from (\ref{Eq_71}) and (\ref{Eq_72}) apply to this case as well as to the combination of a liquid with its vapour or with the corresponding solid body, and only the values of the functions under consideration are different for the individual cases.

For example, the pressure and specific volume of a saturated vapour are represented by different functions of temperature when the vapour is in contact with its liquid than when it is in contact with the corresponding solid body.
\vspace*{-2mm}
\begin{center}
--------------------------------------------------- 
\end{center}
\vspace*{-2mm}

It may happen that for a certain value of $T$ the values of the quantities $v_1$ and $v_2$, as they result from equations (\ref{Eq_71}) and (\ref{Eq_72}), become equal.
Then the two states of matter which are in contact with each other are identical, and the state of equilibrium considered here coincides with that discussed above in the 1st solution, in that the whole body is homogeneous.

Such a value of $T$ is the so-called critical temperature, for which the saturated vapour is identical 
to the liquid in contact.
This value of the temperature, and that of the corresponding specific volume $v_1=v_2$, determine the critical state, which represents a continuous transition from the liquid to the gaseous state, in that an increase in the specific volume (at constant temperature) brings about the gaseous state and a decrease in the same brings about the liquid state.

For lower temperatures the values of $v_1$ and $v_2$ are then different, and for higher temperatures they become imaginary, so that in the latter case the 2nd solution considered here has no sense, and one must go back to the 1st, which corresponds to a homogeneous state of equilibrium.

It is easy to set up the equations from which the determinants of the critical state ($T$ and $v$) can be calculated. 
If we assume $v_1$ and $v_2$ to be infinitesimally different from each other, and by setting $v_1=v+d\,v$ and $v_2=v$, equations (\ref{Eq_71}) and (\ref{Eq_72}) provide the following conditions:

{1) From (\ref{Eq_71}):} 
\vspace*{-3mm}
\begin{align}
  P \;+\; \frac{\partial P}{\partial v} \:.\; dv
  & \:=\;
  P
  \;\: , 
  \nonumber 
\end{align}
therefore: 
\vspace*{-3mm}
\begin{align}
  \frac{\partial P}{\partial v}  
  & \:=\;
  0
  \;\: . 
  \label{Eq_82} 
\end{align}

{2) From (\ref{Eq_72}):} 
\vspace*{-3mm}
\begin{align}
 \bigintsss_{\,v}^{v+dv}\!
     P \:.\; \partial v
  & \:=\;
  P \:.\; dv  
  \;\: , 
  \nonumber 
\end{align}
or: 
\vspace*{-3mm}
\begin{align}
  P \:.\; d v
  \;+\;
  \frac{\partial P}{\partial v}  
  \:.\:
  \frac{dv^2}{1\,.\,2}
  \;+\;
  \frac{\partial^2 P}{\partial\,v^2}  
  \:.\:
  \frac{dv^3}{1\,.\,2\,.\,3}
  & \:=\;
  P \:.\; dv  
  \;\: , 
  \nonumber 
\end{align}
and from this, with regard to (\ref{Eq_82}): 
\vspace*{-3mm}
\begin{align}
  \frac{\partial^2 P}{\partial\,v^2}  
  & \:=\;
  0  
  \;\: . 
  \label{Eq_83} 
\end{align}
The two equations (\ref{Eq_82}) and (\ref{Eq_82}) provide the values of the critical temperature $T$ and the corresponding specific volume $v$ in the critical state, which can therefore be calculated directly if the dependence of the pressure $P$ on $T$ and $v$ is generally known.
\vspace*{-2mm}
\begin{center}
--------------------------------------------------- 
\end{center}
\vspace*{-2mm}

{\bf \underline{Third Solution}.} 
All $3$ parts of the body are different.

  Internal conditions: 
\begin{equation}
\left.
\begin{aligned}
  P_1 & \:=\; P_2  \:=\; P_3 
  \;\: ,  \\
  \bigintsss_{\,v_2}^{v_1}\!\! P\;\partial\,v \:
  & \:=\; P_1 \:.\: \left(\: v_1 \:-\: v_2 \:\right) 
  \;\: ,  \\
  \bigintsss_{\,v_3}^{v_2}\!\! P\;\partial\,v \:
  & \:=\; P_1 \:.\: \left(\: v_2 \:-\: v_3 \:\right) 
  \;\: . 
\end{aligned}
\;\;\;\;\;\;\;\;
\right\} 
\label{Eq_84} 
\end{equation}
These $4$ equations
{\color{red}{\it(with two equations contained in
the first one $P_1\,=\,P_2\,=\,P_3$)}}  
result in very specific values of the $4$ unknowns $T$, $v_1$, $v_2$, $v_3$.

  The external conditions are:
\begin{equation}
\left.
\begin{aligned}
  \sum_k M_k 
  \;\:& \:=\; M
  \;\: ,  \\
  \sum_k M_k \: v_k
  & \:=\; V
  \;\: ,  \\
  \sum_k M_k \: u_k
  & \:=\; U
  \;\: , 
\end{aligned}
\;\;\;\;\;\;\;\;
\right\} 
\label{Eq_85} 
\end{equation}
and from this the values of $M_1$, $M_2$, $M_3$.

This solution corresponds to a state in which all $3$ aggregate states of the body are in contact with each other {\color{red}{\it(vapour, liquid and solid)}}. As you can see, this is only possible for a very specific temperature, which results from the $4$ equations (\ref{Eq_84}) and is almost $= 0$°C for water {\color{red}{\it(\,namely the ``\,triple point\,''\,)}}, for example. This temperature is the temperature for which boiling point and melting point correspond to the same {\color{red}{\it(\,``\,saturation\,''\,)}} pressure.

However, this solution also only has a physical sense if the values of $M_1$, $M_2$, $M_3$ resulting from the external conditions (\ref{Eq_85}) are all positive, and according to (\ref{Eq_70}) it only corresponds to a state of equilibrium if the quantities 
$[\,\partial\,P/\partial\,v\,]_1$, 
$[\,\partial\,P/\partial\,v\,]_2$, 
$[\,\partial\,P/\partial\,v\,]_3$, 
which have very specific values for a particular body, are all negative.
However, if these conditions are fulfilled, the equilibrium is absolutely stable, i.e. the corresponding value of entropy is greater than any other value obtained by the two previous solutions under the same external conditions
(we skip the detailed proof of this theorem here for the same reasons as in the discussion of the 2nd solution). 

An example of this is obtained when water in contact with saturated water vapour is cooled below $0$°C {\color{red}{\it(\,i.e. ``\,supercooled\,'' liquid water\,)}}.

This state corresponding to the 2nd solution is unstable, in that a small external disturbance causes the equilibrium to be permanently abandoned and the equilibrium state corresponding to the 3rd solution to be assumed, in which the various parts of the water at $0$°C are found side by side in solid, liquid and vapour form.  In contrast, before cooling below $0$°C, the state of equilibrium corresponding to the second solution was stable, because in this case the external conditions are such that the values of $M_1$, $M_2$, $M_3$ resulting for the 3rd solution from (\ref{Eq_85}) are not all positive, so this solution has no meaning.

We are now in a position to indicate the general procedure to be followed in order to determine the stable state of equilibrium of a body of which the mass $M$, the volume $v$ and the energy $U$ are given: first examine whether positive values of $M_1$, $M_2$, $M_3$ result from the 3rd solution by means of the $3$ equations (\ref{Eq_85}).

If so, the equilibrium corresponds to this solution, and the body is in $3$ different states of aggregation side by side; but if not, consider the 2nd solution; and if this also makes no sense, the body is completely homogeneous. In terms of the stability of the equilibrium, the 3rd solution therefore has priority over the first two, the 2nd over the first.

To give an example of this, let us imagine $M$, $V$ and $U$ given in such a way that the 3rd solution makes no sense, but the 2nd does, in that the values of $M_1$ and $(M_2+M_3)$ are positive if they are calculated together with $T$ from the $3$ equations (\ref{Eq_73}) which express the external conditions of the 2nd solution. $M_1$, for example, represents (saturated) vapour, and $(M_2+M_3)$ liquid.  Then the equilibrium is given by this solution. If we now leave $M$ constant, while $V$ and $U$ both grow (for example by adding heat) in such a way that $T$ remains constant, then according to (\ref{Eq_71}) and (\ref{Eq_72}) $v_1$ and $v_2$ also become constant, consequently $P1$, $u_1$, $u_2$ will also remain constant, and an insight into the equations (\ref{Eq_73}) teaches that, since $v_1>v_2$ and $u_1>u_2$, $M_1$ will grow, whereas $(M_2 + M_3)$ will decrease.

If the process is continued, $M_1=M$ and $(M_2+M_3)=0$. In this case, the whole body is gaseous and homogeneous, so the 2nd solution coincides with the 1st. 
If you now go even further in the direction described (by adding more heat), $(M_2+M_3)$ becomes negative, the 2nd solution thus loses its meaning, and the stable equilibrium is now represented by the 1st solution, i.e. the vapour remains homogeneous from now on, and the pressure, which was constant until then, begins to change at the same time as the specific volume. This characterises the behaviour of saturated and superheated vapours at a constant temperature.

Similar examples could be made with other states of matter and in relation to the 3rd solution.
\vspace*{-2mm}
\begin{center}
--------------------------------------------------- 
\end{center}
\vspace*{-2mm}

We have based the whole investigation on the assumption that the mass, volume and energy of the body are definitely given, and have considered the maximum entropy corresponding to this assumption. However, the given external conditions can also have a form other than this.  In any case, however, the principle to be applied to find the state of equilibrium remains one and the same : the maxima of entropy of the body possible under the given conditions are determined, and the absolute maximum corresponds to the absolutely stable equilibrium, whereas any other maximum corresponds to a more or less unstable state of equilibrium.

A particularly common case is, for example, that in addition to the mass, the pressure $P$ under which the body is located is given, and also the energy of the body for any volume, which always changes when the body changes its volume, because external work is then performed.

We also want to set up the equations for this case, from which the determinants of the equilibrium are calculated. If the entropy is to reach a maximum for a state, all possible infinitely small variations of the entropy in this state must be $=0$.
We therefore obtain, as in (\ref{Eq_64}) {\it\color{red}--\,in fact as in (\ref{Eq_60})?\,--\,}, if we retain all the terms used there, the condition: 
\vspace*{0mm}
\begin{align}
 \delta\,S & 
  \;=\; \sum_k \: \frac{M_k \:.\: \delta\,u_k}{T_k}
  \;+\; \sum_k \: \frac{M_k \:.\: P_k\:.\: \delta\,v_k}{T_k}
  \;+\; \sum_k \: s_k \:.\: \delta\,M_k
  \;=\; 0
\; . \label{Eq_86} 
\end{align}
\vspace*{-5mm}

According to equations (\ref{Eq_61}), (\ref{Eq_62}) and (\ref{Eq_63}), the following external conditions apply: \vspace*{2mm} \\
{\bf -- Since the mass $M$} of the body remains constant: 
\vspace*{0mm}
\begin{align}
  \sum_k \: \delta\,M_k  & \;=\; 0
\; . \label{Eq_87} 
\end{align}
\vspace*{-3mm} \\
{\bf -- Since the energy $U$} changes according to the external work, which is determined by the given external pressure $P$ and by the variation of the total volume $V$: 
\vspace*{0mm}
\begin{align}
 \delta\,U \;+\; P \:.\; \delta\,V  \;=\; 0 \; , 
 \hspace*{25mm}
 \hspace*{25mm}
 \nonumber \\
\mbox{or:} \;\;\;
   \delta \left(
   \sum_k \: M_k \:.\: u_k
   \right)
  \;+\; 
   P \:.\; \delta \left(
  \sum_k \: M_k \:.\: v_k
    \right)
 \;=\; 0 \; , 
 \hspace*{25mm}
 \nonumber \\
\mbox{or:} \;\;\;
  \sum_k \: M_k \:.\: \delta\,u_k
  \;+\; 
  \sum_k \: u_k \:.\: \delta M_k
  \;+\; 
  P\:.\: \sum_k \: M_k \:.\: \delta\,v_k
  \;+\; 
  P\:.\: \sum_k \: v_k \:.\: \delta M_k
  \;=\; 0
\; . \label{Eq_88} 
\end{align}
\noindent Then, using the two equations (\ref{Eq_87}) and (\ref{Eq_88}), we can eliminate any 2 variations from (\ref{Eq_86}), e.g. $\delta M_2$ and $\delta\,u_2$, so that we then only obtain completely independent variations in the expression of $\delta S$. After performing the calculation, the result is: 
\vspace*{0mm}
\begin{align}
\delta\,S  & \:=\; 
 \left\{ 
   \frac{1}{T_1} \:-\: \frac{1}{T_2} 
 \right\} .\: M_1 \:.\: \delta\,u_1
 \;-\;
 \left\{ 
   \frac{1}{T_2} \:-\: \frac{1}{T_3} 
 \right\} .\: M_3 \:.\: \delta\,u_3
\nonumber \\
& \;+\;
 \left\{ 
   \frac{P_1}{T_1} \:-\: \frac{P}{T_2} 
 \right\} .\: M_1 \:.\: \delta\,v_1
 \;+\;
 \left\{ 
   \frac{P_2}{T_2} \:-\: \frac{P}{T_2} 
 \right\} .\: M_2 \:.\: \delta\,v_2
 \;+\;
 \left\{ 
   \frac{P_3}{T_3} \:-\: \frac{P}{T_2} 
 \right\} .\: M_3 \:.\: \delta\,v_3
\; . \nonumber 
\nonumber \\
& \;+\;
 \left\{ (s_1\:-\:s_2)
   \:-\frac{(u_1\:-\:u_2)\:+\:P\:(v_1\:-\:v_2)}{T_2}
 \right\} .\: \delta M_1
\nonumber \\
& \;-\;
 \left\{ (s_2\:-\:s_3)
   \:-\frac{(u_2\:-\:u_3)\:+\:P\:(v_2\:-\:v_3)}{T_2}
 \right\} .\: \delta M_3
 \:\;=\:\; 0 
 \; , \nonumber %
\end{align}
from which the following equations emerge if the coefficients of the $7$ independent variations {\it\color{red}(all terms in curly brackets in the form ``\,$\{...\}\:\delta ...$\,''\,)} are set to 
$0$:$\,$\footnote{$\:${\color{red}I have introduced the common temperature $T$, where Planck used $T_1$ in the last two equations. (P. Marquet)}}
\vspace*{0mm}
\begin{align}
T_1 \;=\; T_2 & \:=\; T_3 {\color{red}\;\;=\; T} \; ,
\nonumber \\
P_1 \;=\; P_2 & \:=\; P_3 \;=\; P \; ,
\nonumber \\
s_1\:-\:s_2 & \:=\;
 \frac{(u_1\:-\:u_2)\:+\:P\:(v_1\:-\:v_2)}{{\color{red}T}} \; ,
\nonumber \\
s_2\:-\:s_3 & \:=\;
 \frac{(u_2\:-\:u_3)\:+\:P\:(v_2\:-\:v_3)}{{\color{red}T}} \; .
\nonumber
\end{align}
If we compare this result with that obtained under (\ref{Eq_65}), (\ref{Eq_66}), (\ref{Eq_67}) and (\ref{Eq_68}), we immediately recognise that the internal conditions of equilibrium, i.e. those conditions which are dependent only on the nature of the body under  
consideration,\footnote{$\:${\color{red}We recognize here the fundamental aim of Max Planck and his general seek for  \dashuline{\,absolute physical relationship\,}, namely independent of special features like the kind of coordinate systems, or here like the number or kind of bodies. (P. Marquet)}}
but not on the given external circumstances in which it finds itself, are the same in both cases, as this is in the nature of things.

In order to obtain the external equilibrium conditions, the given mass, the given pressure and the given energy (for a given volume) must be taken into account. 

Here too, it is possible to distinguish between $3$ different types of solutions, depending on whether the body is completely homogeneous or whether it is in $2$ or $3$ different states of aggregation. The relevant investigations can be carried out in the same way as in the case above. Here too, as far as the stability of the equilibrium is concerned (i.e. the third solution, if it makes sense) has priority over the first two, the second over the first.

An example of the latter circumstance is given by water when it is cooled below $0$°C under the constant pressure of the atmosphere. The equilibrium then corresponds to the 1st solution. However, while the equilibrium was stable before cooling, because the 2nd solution had no meaning there, it is unstable after cooling because the 2nd solution now gives positive values of $M_1$ and $(M_2+M_3)$. Therefore, if the equilibrium is slightly disturbed, a new equilibrium state will occur which corresponds to the 2nd solution, i.e. a combination of ice and water.

Whatever the circumstances to which the body is subjected may be, the internal conditions of equilibrium, as expressed in equations (\ref{Eq_69}) and grouped into the $3$ different solutions discussed above, always remain the same and serve as a basis for finding the equilibrium once and for all. To these are then added the external conditions, which are derived from the specific circumstances given in each case and fully determine the state of equilibrium of the body. Of course, this presupposes that $P$ and $u$ are generally known as functions of $T$ and $v$, which is not yet the case.

%
%
%

%
%
%

%
%
%

%
%
%

%
%
%

%
%
%


%
%

%


\vspace*{-2mm}
\begin{center}
--------------------------------------------------- 
\end{center}
\vspace*{-2mm}

\noindent
Kgl. Hof- und UniversitätsLuchdruckerei von Dr. C. Wolf \& Sohn.
\\
Royal Court and University Printing House of Dr C. Wolf \& Son.

\end{document}